\documentclass{elsart}

\usepackage{graphicx}

\usepackage{amssymb,amsbsy}

\begin{document}



\newcommand{\Ha}{\mbox{\sl Ha}}
\newcommand{\Ma}{\mbox{\sl Ma}}
\newcommand{\Reyn}{\mbox{\sl Re}}
\newcommand{\Pran}{\mbox{\sl Pr}}
\newcommand{\Al}{\mbox{\sl Al}}
\newcommand{\eff}{\mathit eff}
\newcommand{\diag}{\mbox{diag}}

\begin{frontmatter}



\title{Steady State Convergence Acceleration of the Generalized Lattice Boltzmann Equation with Forcing Term through Preconditioning}

\author[label1,label2]{Kannan N. Premnath\corauthref{cor1}}
\ead{nandha@metah.com}
\author[label1]{Martin J. Pattison}
\author[label1,label2,label3,label4]{Sanjoy Banerjee}
\address[label1]{MetaHeuristics LLC, 3944 State Street, Suite 350, Santa Barbara, CA 93105}
\address[label2]{Department of Chemical Engineering, University of California Santa Barbara, Santa Barbara, CA 93106}
\address[label3]{Department of Mechanical Engineering, University of California Santa Barbara, Santa Barbara, CA 93106}
\address[label4]{Bren School of Environmental Science and Management, University of California Santa Barbara, Santa Barbara, CA 93106}
\corauth[cor1]{Corresponding author.}

\author{}

\begin{abstract}
Several applications exist in which lattice Boltzmann methods (LBM)
are used to compute stationary states of fluid motions, particularly
those driven or modulated by external forces. Standard LBM, being
explicit time-marching in nature, requires a long time to attain
steady state convergence, particularly at low Mach numbers due to
the disparity in characteristic speeds of propagation of different
quantities. In this paper, we present a preconditioned generalized
lattice Boltzmann equation (GLBE) with forcing term to accelerate
steady state convergence to flows driven by external forces. The  use of
multiple relaxation times in the GLBE allows enhancement of the numerical stability.
Particular focus is given in preconditioning external
forces, which can be spatially and temporally dependent. In
particular, correct forms of moment-projections of source/forcing
terms are derived such that they recover preconditioned
Navier-Stokes equations with non-uniform external forces. As an
illustration, we solve an extended system with a preconditioned lattice kinetic
equation for magnetic induction field at low magnetic Prandtl
numbers, which imposes Lorentz forces on the flow of conducting
fluids. Computational studies, particularly in three-dimensions, for
canonical problems show that the number of time steps needed to reach
steady state is reduced by orders of magnitude with preconditioning.
In addition, the preconditioning approach resulted in significantly improved
stability characteristics when compared with the corresponding single
relaxation time formulation.
\end{abstract}

\end{frontmatter}

\section{Introduction} \label{section_introduction}
In recent years, the lattice Boltzmann method (LBM) has emerged as
an alternative and accurate approach for computational physics, and,
in particular, for computational fluid dynamics (CFD) problems
~\cite{chen98,succi01}. It is generally based on minimal discrete
kinetic models whose emergent behavior, under appropriate
constraints, corresponds to the dynamical equations of fluid flows
or other physical systems. It involves the solution of the
lattice-Boltzmann equation (LBE) that represents the evolution of
the distribution of particle populations due to their collisions and
advection on a lattice. When the lattice, which represents the
discrete directions for propagation of particle populations, satisfies sufficient
rotational symmetries, the LBE recovers the weakly compressible
Navier-Stokes equations (NSE) in the continuum limit. The LBE can be
constructed to simulate complex flows by incorporating additional
physical models~\cite{shan93,swift96}.

Though it evolved as a computationally efficient form of lattice gas
cellular automata~\cite{higuera89}, it was well established about a
decade ago that the LBE is actually a much simplified form of the
continuous Boltzmann equation~\cite{he97,abe97}. As a result, several previous
results in discrete kinetic theory could be directly applied to the
LBE. This led to, for example, improved physical modeling in
various situations, such as multiphase flows~\cite{luo00,he02} and
multicomponent flows~\cite{asinari06}, and in an asymptotic theory suitable
for rigorous numerical analysis~\cite{junk05}. In particular, the latter development
has made it possible to study consistency, convergence and accuracy
of the LBE in a manner similar to the classical numerical methods for continuum based
approaches. As a result of features of the stream-and-collide
procedure of the LBE such as the algorithmic simplicity, amenability
to parallelization with near-linear scalability, and its ability to
represent complex boundary conditions and incorporate physical
models more naturally, it has rapidly found a wide range of
applications~\cite{ladd01,succi02,yu03,nourgaliev03,premnath05}.

Several applications exist where steady state solutions to fluid
flow problems are highly desirable. Examples include
magnetohydrodynamic flows and multiphase porous media, where one is
often mainly interested in investigations of their steady state
characteristics. On the other hand, the standard form of the LBE is
hyperbolic in nature and its solution involves explicit marching in
time. As a result, it necessarily involves evolving through a
transient phase before reaching a stationary state. Due to the need
to march for many number of time steps in this transient phase,
it incurs significant computational cost. Another
important related issue is that the LBE actually represents
compressible NSE valid at low Mach numbers, $\Ma$. Its deviation from
the incompressible NSE, which we shall call ``compressibility
deviations", is independent of grid resolution. When one intends to
simulate close to incompressible flow using LBE, such deviations (or
the $\Ma$) should be made smaller. This is also desirable from a
computational viewpoint as the stability regime of the LBE generally
widens at lower $\Ma$. However, as the $\Ma$ is lowered, there is a
greater disparity between the propagation speeds of density
perturbations, i.e. the speed of sound, and the convection speed of
the fluid. In a hyperbolic system, the numerical domain of influence
should encompass the physical domain~\cite{leveque02}, requiring
resolution of the time scales of the fluid motion. As a result,
computing lower $\Ma$ flows further compounds the issue and requires
a larger number of time steps to achieve steady state convergence.

In recent years, several approaches have been proposed to improve the
convergence rate of the LBE to steady state. These include, in one
category, reformulations of the LBE to time-independent versions
that can be solved as a linear system~\cite{verberg99,bernaschi01} and
a finite-difference time-independent version solved by a multigrid
method~\cite{tolke02}. In another, they involve adding an artificial
body force to the time-dependent LBE~\cite{kandhai01},
constructing an implicit LBE in a finite-difference or
finite-element formulation that allows taking larger time
steps~\cite{lee03,seta02,tolke98}, or by using a non-linear form of
multigrid solver with a non-linear LBE time stepping
scheme~\cite{mavriplis06}. All these schemes can significantly
improve convergence rates, but at the cost of increased complexity
as compared with the standard LBE.

On the other hand, Guo \emph{et al}.~\cite{guo04} proposed an
alternate approach to reduce the number of time steps necessary for steady state
convergence by applying preconditioning to the LBE, while maintaining its
simplicity. The essential principle of this approach, which was originally
developed for general hyperbolic schemes by Turkel and others, is as
follows~\cite{turkel87,vanleer91,choi93,turkel99}. At low $\Ma$, in
explicit formulations, there is a disparity in propagation speeds of
density perturbation and fluid convection. This is formally
characterized by higher values of condition number, which is defined
as the ratio of the fastest to the slowest speeds of propagation, or
equivalently, the ratio of the maximum to minimum eigenvalues of the
hyperbolic system, and is inversely proportional to $\Ma$. By
applying a preconditioner, the speeds of propagation of various
quantities can be made closer to one another. This can be achieved
only at the cost of sacrificing the temporal accuracy of the
solutions, which in any case is not very important as the chief
interest is in obtaining steady state flow characteristics. Guo
\emph{et al}.~\cite{guo04} achieved this in the context of
LBE by applying a preconditioning parameter that modifies the
equilibrium distribution function in its collision model. Its
emergent behavior is a preconditioned compressible NSE with reduced
stiffness and hence significantly reduces the number of time steps to reach
steady state.

All the preconditioning approaches for LBM mentioned above employ
the single relaxation time (SRT) model~\cite{bhatnagar54} to
represent the effect of particle collisions, with the exception of a recent
work that adopts a different approach to preconditioning a general form of the
LBE than considered here~\cite{izquierdo08}. A commonly used form,
the SRT-LBE involves relaxation of particle distributions to their
local equilibria at a rate determined by a single
parameter~\cite{qian92,chen92}. On the other hand, an equivalent
representation of distribution functions is in terms of their
moments, such as various hydrodynamic fields including density, mass
flux, and stress tensor. The relaxation process due to collisions
can more naturally be described in terms of a space spanned by such
moments, which can in general relax at different rates. This forms
the basis of the generalized lattice-Boltzmann equation (GLBE) based
on multiple relaxation times
(MRT)~\cite{dhumieres92,lallemand00,dhumieres02}. By carefully
separating the time scales of various hydrodynamic and kinetic modes
through a linear stability analysis, the numerical stability of the
GLBE or MRT-LBE can be significantly improved when compared with the
SRT-LBE, particularly for more demanding problems at high Reynolds
numbers~\cite{lallemand00}. The MRT-LBE has also been extended for
multiphase
flows~\cite{mccracken05,premnath06,premnath05a,mccracken05a,tolke06}, and,
more recently, for magnetohydrodynamic problems~\cite{pattison07},
with superior stability characteristics. It has also been used for
LES of a class of turbulent flows~\cite{krafczyk03,yu06,premnath08}.
It is known that for a given grid resolution and Reynolds number,
the standard LBM based on the SRT model becomes less stable as $\Ma$
is lowered due to the relaxation time becoming
smaller~\cite{he97d,guo04}. Since the preconditioning is mainly
intended to accelerate steady state convergence at lower $\Ma$, it is
also important to stabilize the computations, which can be optimally
achieved by using the MRT-LBE.

Another consideration is how to precondition the LBE in the presence
of external forces. While Guo \emph{et al}.~\cite{guo04}
suggest a way to precondition a particular form of forcing term,
details on preconditioning general forms of spatially and temporally
varying forcing terms are lacking. Such forms are important in many
situations including magnetohydrodynamic (MHD) flows, where the
Lorentz force impressed on the fluid can vary spatially and
temporally, and multiphase flows represented by mean-field models,
and buoyant flows. Moreover, previous studies were limited to a
narrower class of two-dimensional (2D) flows, largely in the absence
of any body force. In addition, dynamics of flow of complex fluids
is generally represented by a system of LBE, typically with one LBE
representing the flow fields and another one characterizing the
evolution of other physical processes occurring within the fluid.
For example, for MHD flow, we have one LBE to represent the fluid
flow and another one for the magnetic induction equation. Similarly,
in the case of multiphase flows, we have two sets of LBE -- one for
the fluid dynamics and another one for the dynamics of an order
parameter that distinguishes the phases. Thus, it is also important
to extend the preconditioning to such systems of LBE.

It is important to note that preconditioning a system of LBE formally improves
the condition number of its equivalent macroscopic system. For example, in the
context of MHD, preconditioning a system of LBE actually improves the
condition number of the equivalent system consisting of the NSE and
the magnetic induction equation. As a result, while the convergence rate of the
LBE scheme, which is typically associated with an exponent, is unchanged, the
prefactor of the convergence rate is modified by preconditioning.
In effect, the number of time steps needed to reach a steady state representation
of the equivalent macroscopic system is significantly reduced.

The primary objective of this paper is then to develop a
preconditioning method for the MRT-LBE with general forms of forcing
terms representing non-uniform forces to accelerate convergence to steady
state flows. In this regard, we derive expressions for preconditioned equilibrium
moments that gives rise to the linear viscous and non-linear convective
behavior of a fluid. Based on a Chapman-Enskog multiscale
analysis~\cite{chapman64}, we also derive correct functional forms
of the moment projections of source/forcing terms corresponding to spatially
and temporally dependent variation of forces, which avoids discrete
lattice artifacts. A limiting case of the source terms for the
SRT-LBE will also be presented. To illustrate the use of
preconditioning for a system of LBEs, we derive a preconditioned
lattice kinetic model for MHD, and also provide a simple
approach to attain lower values of magnetic Prandtl number at steady
state, which is important for simulating liquid metal flows. We
illustrate the advantages of these approaches for a set of canonical
problems, particularly in three-dimensions (3D). In doing so, we
also present some new results with shear driven MHD flows. It may be
noted that the approach presented here, though illustrated for MHD
problems, may be readily extended to develop preconditioning to a
system of MRT-LBEs for a variety of other problems.

This paper is organized as follows. After a brief description of the
generalized lattice-Boltzmann equation with forcing term in
Sec.~\ref{subsec:glbe}, in Sec.~\ref{subsec:pglbe} we present a
derivation of the preconditioned GLBE with forcing term in both 2D
and 3D. The corresponding preconditioned form of lattice kinetic
equation for magnetic induction is discussed in
Sec.~\ref{sec:pvlbe}. Some canonical examples simulated using
preconditioned LBM are discussed in Sec.~\ref{sec:results}. Finally,
the summary and conclusions of this paper are provided in
Sec.~\ref{sec:summary}.

\section{\label{subsec:glbe}Generalized Lattice Boltzmann Equation with Forcing Term}
The lattice-Boltzmann method computes the evolution of distribution
functions as they move and collide on a lattice grid. The collision
process considers their relaxation to their local equilibrium
values, and the streaming process describes their movement along the
characteristics directions given by a discrete particle velocity
space represented by a lattice. Typical lattice velocity models
include the two-dimensional, nine velocity (D2Q9) and the
three-dimensional, nineteen velocity (D3Q19) models~\cite{qian92},
which are considered in this paper. The particle velocity
$\overrightarrow{e_{\alpha}}$ corresponding to the D2Q9
model may be written as:
\begin{equation}
\overrightarrow{e_{\alpha}} = \left\{\begin{array}{ll}
   {(0,0)}&{\alpha=0}\\
   {(\pm 1,0),(0,\pm 1)}&{\alpha=1,\cdots,4}\\
   {(\pm 1,\pm 1)} &{\alpha=5,\cdots,8}
\end{array} \right.
\label{eq:velocityd2q9}
\end{equation}
and for the D3Q19 model:
\begin{equation}
\overrightarrow{e_{\alpha}} = \left\{\begin{array}{ll}
   {(0,0,0)}&{ \alpha=0}\\
   {(\pm 1,0,0),(0,\pm 1,0),(0,0,\pm 1)}&{ \alpha=1,\cdots,6}\\
   {(\pm 1,\pm 1,0),(\pm 1,0,\pm 1),(0,\pm 1,\pm 1)}&{ \alpha=7,\cdots,18}.
\end{array} \right.
\label{eq:velocityd3q19}
\end{equation}

The GLBE computes collisions in moment space, while the streaming
process is performed in the usual particle velocity
space~\cite{dhumieres02}. The form of the GLBE considered
here~\cite{premnath08} also computes the forcing term, which
represents the effect of external forces as a second-order accurate
time-discretization, in moment space~\cite{premnath06,premnath08}.
We use the following notation in our description of the procedure
below: In \emph{particle velocity space}, the local distribution
function $\mathbf{f}$, its local equilibrium distribution
$\mathbf{f}^{eq}$, and the source terms due to external forces
$\mathbf{S}$ may be written as the following column vectors:
$\mathbf{f}=\left[ f_0,f_1,f_2,\ldots,f_{b} \right]^{\dag}$,
$\mathbf{f}^{eq}=\left[ f_0^{eq},f_1^{eq},f_2^{eq},\ldots,f_{b}^{eq}
\right]^{\dag}$, and $ \mathbf{S}=\left[ S_0,S_1,S_2,\ldots,S_{b}
\right]^{\dag}$, where $b$ is the number of non-zero discrete
velocity directions for a given lattice model. Thus, $b=8$ and
$b=18$ for D2Q9 and D3Q19 models, respectively. Here, the
superscript $\dag$ represents the transpose operator.

In particular, the form of the source terms in particle velocity
space are obtained from the expression used in the discrete velocity
Boltzmann equation
$-\overrightarrow{F}/\rho\cdot\overrightarrow{\nabla}_{\overrightarrow{e}_{\alpha}}f_{\alpha}$
by approximating it to
$-\overrightarrow{F}/\rho\cdot\overrightarrow{\nabla}_{\overrightarrow{e}_{\alpha}}f_{\alpha}^{eq,M}$
~\cite{he98} and further simplifying by neglecting terms of the
order of $O(\Ma^2)$ or higher to get~\cite{premnath06} $ S_{\alpha} =
w_{\alpha}\left[
3\left(\overrightarrow{e}_{\alpha}-\overrightarrow{u}\right)+
            9\left(  \overrightarrow{e_{\alpha}} \cdot  \overrightarrow{u} \right)\overrightarrow{e}_{\alpha
            }\right]\cdot\overrightarrow{F}
$ where
$f_{\alpha}^{eq,M}=w_{\alpha}\{1+3\overrightarrow{e_{\alpha}}\cdot\overrightarrow{u}+9/2(\overrightarrow{e_{\alpha}}\cdot\overrightarrow{u})^2-1/2\overrightarrow{u}^2\}$
is the local discrete Maxwellian truncated to
$O(\Ma^2)$~\cite{qian92}. Here, $w_\alpha$ is a weighting
factor~\cite{qian92}, $\rho$ and $\overrightarrow{u}$ are the local
fluid density and velocity, respectively, and $\overrightarrow{F}$ is the
external force field, whose Cartesian components are $F_x$, $F_y$ and $F_z$.

The moments $\mathbf{\widehat{f}}$ are related to the distribution
function $\mathbf{f}$ through the relation
$\mathbf{\widehat{f}}=\mathcal{T}\mathbf{f}$ where $\mathcal{T}$ is
the transformation matrix. Here, and in the following, the ``hat"
represents the moment space. The transformation matrix $\mathcal{T}$
is constructed such that the collision matrix in moment space
$\widehat{\Lambda}$ is a diagonal matrix through
$\widehat{\Lambda}=\mathcal{T}\Lambda\mathcal{T}^{-1}$, where
$\Lambda$ is the collision matrix in particle velocity space. The
elements of $\mathcal{T}$ are obtained in a suitable orthogonal
basis as combinations of monomials of the Cartesian components of
the particle velocity $\overrightarrow{e_{\alpha}}$ through the
standard Gram-Schmidt procedure, which are provided by Lallemand and
Luo~\cite{lallemand00} and d'Humi{\`e}res \emph{et al}.~\cite{dhumieres02} for 2D and 3D lattice models,
respectively. Similarly, the equilibrium moments and the source
terms in moment space may be obtained through the transformation
$\mathbf{\widehat{f}}^{eq}=\mathcal{T}\mathbf{f}^{eq}$,
$\mathbf{\widehat{S}}=\mathcal{T}\mathbf{S}$. The components of
moment-projections of these quantities are:
$\mathbf{\widehat{f}}=\left[
\widehat{f}_0,\widehat{f}_1,\widehat{f}_2,\ldots,\widehat{f}_{b}
\right]^{\dag}$, $\mathbf{\widehat{f}}^{eq}=\left[
\widehat{f}_0^{eq},\widehat{f}_1^{eq},\widehat{f}_2^{eq},\ldots,\widehat{f}_{b}^{eq}
\right]^{\dag}$, and $ \mathbf{\widehat{S}}=\left[
\widehat{S}_0,\widehat{S}_1,\widehat{S}_2,\ldots,\widehat{S}_{b}
\right]^{\dag}$. The expressions for these quantities are provided
in Appendix~\ref{app:momentcomponents} for both D2Q9 and D3Q19
models.

The solution of the GLBE with forcing term can be written in terms
of the following ``effective" collision and streaming steps,
respectively:
\begin{equation}
\mathbf{\widetilde{f}}(\overrightarrow{x},t)=\mathbf{f}(\overrightarrow{x},t)+\boldsymbol{\varpi}(\overrightarrow{x},t),
\label{eq:postcollision}
\end{equation}
and
\begin{equation}
f_{\alpha}(\overrightarrow{x}+\overrightarrow{e}_{\alpha}\delta_t,t+\delta_t)=\widetilde{f}_{\alpha}(\overrightarrow{x},t),
\label{eq:streaming}
\end{equation}
where the distribution function
$\mathbf{f}=\{f_{\alpha}\}_{\alpha=0,1,\ldots,b}$ is updated due to
``effective" collisions resulting in the post-collision distribution
function
$\mathbf{\widetilde{f}}=\{\widetilde{f}_{\alpha}\}_{\alpha=0,1,\ldots,b}$
before being shifted along the characteristic directions during the
streaming step. The change in distribution function due to
collisions as a relaxation process and external forces is
represented by $\boldsymbol{\varpi}$, and following Premnath
\emph{et al}.~\cite{premnath08} it can written as
\begin{equation}
\boldsymbol{\varpi}(\overrightarrow{x},t)= \mathcal{T}^{-1}\left[
-\widehat{\Lambda}\left(\mathbf{\widehat{f}}-\mathbf{\widehat{f}}^{eq}
\right)+\left(
\mathcal{I}-\frac{1}{2}\widehat{\Lambda}
\right)\mathbf{\widehat{S}} \right],
\label{eq:relax_term}
\end{equation}
where $\mathcal{I}$ is the identity matrix and $
\widehat{\Lambda}=\diag(s_0,s_1,\ldots,s_{b})$ is the diagonal
collision matrix in moment space. Also, here and henceforth,
$\mathbf{\widehat{f}}\equiv \mathbf{\widehat{f}}{(\overrightarrow{x},t)}$,
$\mathbf{\widehat{f}}^{eq}\equiv \mathbf{\widehat{f}}^{eq}{(\overrightarrow{x},t)}$ and
$\mathbf{\widehat{S}}\equiv \mathbf{\widehat{S}}{(\overrightarrow{x},t)}$.

It may be noted that Eqs.~(\ref{eq:postcollision}) and (\ref{eq:streaming}) are obtained
from the second-order trapezoidal discretization of the source term
in the GLBE~\cite{premnath06}, viz.,
$f_{\alpha}(\overrightarrow{x} + \overrightarrow{e_{\alpha}}\delta_t,t+\delta_t) - f_{\alpha}(\overrightarrow{x},t) =
-\sum_{\beta}\Lambda_{\alpha \beta} \left[ f_{\beta}(\overrightarrow{x},t) - f_{\beta}^{eq}(\overrightarrow{x},t)
\right] + \varphi_{\alpha}$ where $\varphi_{\alpha}=1/2\left[S_{\alpha
}(\overrightarrow{x},t)+S_{\alpha}(\overrightarrow{x}+\overrightarrow{e_{\alpha}}\delta_t,t+\delta_t)
\right]\delta_t$, which is made effectively time-explicit through a
transformation $\overline{f}_\alpha=f_\alpha-1/2S_\alpha\delta_t$~\cite{he98}, and
then dropping the ``overbar" in subsequent representations for
convenience. Subsequently, both the collision and source terms are
represented in the natural moment space of GLBE. The second-order
discretization provides a more accurate treatment of source terms,
particulary in correctly recovering general forms of external forces
in the continuum limit without spurious terms due to discrete
lattice effects~\cite{guo02}, and its time-explicit representation
faciliates numerical solution in a manner analogous to the standard
LBE.

Now, some of the relaxation times $s_{\alpha}$ in the collision
matrix, i.e. those corresponding to hydrodynamic modes can be
related to the transport coefficients, such as the bulk and shear
viscosities. The rest of the relaxation parameters, i.e. for the
kinetic modes, can be chosen through a von Neumann stability
analysis of the linearized GLBE~\cite{lallemand00,dhumieres02}. See
also Appendix~\ref{app:momentcomponents} for more details.

Once the distribution function is known, the hydrodynamic fields,
i.e., the density $\rho$, velocity $\overrightarrow{u}$, and
pressure $p$ can be obtained as follows:
\begin{equation}
\rho =\sum_{\alpha=0}^{b} f_{\alpha}, \quad \overrightarrow{j}\equiv
\rho\overrightarrow{u} =\sum_{\alpha=0}^{b}
f_{\alpha}\overrightarrow{e}_{\alpha}+\frac{1}{2}\overrightarrow{F}\delta_t,
\quad p=c_s^2\rho,
\end{equation}
where, $c_s=c/\sqrt{3}$ with $c=\delta_x/\delta_t$ being the
particle speed, and $\delta_x$ and $\delta_t$ are the lattice
spacing and time step, respectively.

The computational procedure for the solution of the GLBE with
forcing term is optimized by fully exploiting the special properties
of the transformation matrix $\mathcal{T}$: these include its
orthogonality, entries with many zero elements, and entries with
many common elements that are integers, which are used to form the
most common sub-expressions for transformation between spaces in
avoiding direct matrix multiplications~\cite{dhumieres02}. For
details, we refer the reader to Ref.~\cite{premnath08}. As a result
of such optimizations, the additional computational overhead when
GLBE is used in lieu of the popular SRT-LBE is small, typically
between $15\%-30\%$, but with much enhanced numerical stability that
allows maintaining solution fidelity on coarser grids and also in
simulating flows at higher Reynolds numbers.

\section{\label{subsec:pglbe}Preconditioned Generalized Lattice Boltzmann Equation with Forcing Term for Fluid Flow}
As noted earlier, computation of flows at low $\Ma$ using the
standard LBE can be slow to converge to steady state due to the
condition number of its equivalent NSE being large, as it is inversely
proportional to $\Ma$. Moreover, for a given Reynolds number, there is a limit on how
low $\Ma$ can be before numerical stability problems result, as the relaxation time in the
standard LBE, $\tau$, can become very close to $0.5$ when $\Ma$ is made smaller.
Preconditioning effectively reduces the disparity in propagation speeds of density
perturbation and fluid convection, or improves the condition number
of the equivalent NSE being simulated. The use of the GLBE or MRT-LBE
improves numerical stability by appropriately tuning the relaxation
times of the non-hydrodynamic kinetic or ghost modes through a von
Neumann stability analysis. We now present the preconditioned
generalized lattice Boltzmann equation with forcing term.

Several factors need to be considered in preconditioning the GLBE.
The streaming step in the GLBE is a Lagrangian free-flight or
propagation process from one lattice node to another node. The
collision process is a relaxation step that contains linear, faster
density propagation process and slower viscous momentum transfer
process, and non-linear fluid convective process. They are
individually described in moment space and their separate effects or
contributions need to be properly preconditioned. Also, careful
consideration should be given to the preconditioning of the forcing
terms in moment space, as their contributions, depending on the
moment, vary widely, from simple Cartesian component of external
forces to work due to such forces. In particular, as noted in
Appendix~\ref{app:momentcomponents}, the moment projections of forcing
terms are functions of external force fields and velocities, and their
products. Hence, care needs to be exercised in properly
preconditioning individual components of the forcing terms
corresponding to hydrodynamic and kinetic or ghost modes. As in
Guo \emph{et al}.~\cite{guo04}, we introduce a preconditioning parameter
$\gamma$, with $0 < \gamma \leq 1$, on the GLBE with forcing term.
It may be noted that setting $\gamma$ equal to $1$ reduces to the
standard form without preconditioning, while $\gamma<1$ improves the
condition number of the equivalent NSE system of the GLBE.
By performing a Chapman-Enskog analysis on such GLBE, its preconditioning
can be properly constructed such that it recovers the corresponding preconditioned
compressible NSE in the continuum limit. The details of this procedure carried out
for the D2Q9 model is presented in Appendix~\ref{app:Chapman_Enskog_pGLBE},
which can be extended to other lattice models.

The preconditioned GLBE with forcing term can be written in terms of
the following ``effective" collision and streaming steps,
respectively:
\begin{equation}
\mathbf{\widetilde{f}}(\overrightarrow{x},t)=\mathbf{f}(\overrightarrow{x},t)+\boldsymbol{\varpi}^{*}(\overrightarrow{x},t),
\label{eq:ppostcollision}
\end{equation}
and
\begin{equation}
f_{\alpha}(\overrightarrow{x}+\overrightarrow{e}_{\alpha}\delta_t,t+\delta_t)=\widetilde{f}_{\alpha}(\overrightarrow{x},t),
\label{eq:pstreaming}
\end{equation}
where $\boldsymbol{\varpi}^{*}$ represents the change in
distribution function due to preconditioned collisional relaxation
and forcing terms due to external forces. It can be written as
\begin{equation}
\boldsymbol{\varpi^{*}}(\overrightarrow{x},t)=
\mathcal{T}^{-1}\left[
-\widehat{\Lambda}^{*}\left(\mathbf{\widehat{f}}-\mathbf{\widehat{f}}^{eq,*}
\right)+\left(
\mathcal{I}-\frac{1}{2}\widehat{\Lambda}^{*}
\right)\mathbf{\widehat{S}^{*}} \right].
\label{eq:prelax_term}
\end{equation}
Here, $\mathcal{I}$ is the identity matrix, $ \widehat{\Lambda}^{*}$
is the preconditioned diagonal collision matrix in moment space,
$\mathbf{\widehat{f}}^{eq,*}$ is the preconditioned equilibrium
moments and $\mathbf{\widehat{S}^{*}}$ is the preconditioned moment
projections of source terms due to external forces. Here, and in the
following, the superscript ``$*$" denotes preconditioned variables.

The preconditioning of the components of the equilibrium moments,
\begin{equation}
\mathbf{\widehat{f}}^{eq,*}=\left[
\widehat{f}_0^{eq,*},\widehat{f}_1^{eq,*},\widehat{f}_2^{eq,*},\ldots,\widehat{f}_{b}^{eq,*}
\right]^{\dag}
\end{equation}
which are functions of the conserved moments, can be performed by
analyzing the GLBE using the Chapman-Enksog expansion, as in
Appendix~\ref{app:Chapman_Enskog_pGLBE}. The components of
$\mathbf{\widehat{f}}^{eq,*}$ can be written for the D2Q9 model as:
$\widehat{f}_0^{eq,*} = \rho, \widehat{f}_1^{eq,*} \equiv
e^{eq,*}=-2\rho+3\frac{\overrightarrow{j}\cdot\overrightarrow{j}}{\gamma\rho},
\widehat{f}_2^{eq,*} \equiv
e^{2,eq,*}=\rho-3\frac{\overrightarrow{j}\cdot\overrightarrow{j}}{\gamma\rho},
\widehat{f}_3^{eq,*} = j_x,\widehat{f}_4^{eq,*} \equiv
q_x^{eq,*}=-j_x,\widehat{f}_5^{eq,*} = j_y, \widehat{f}_6^{eq,*}
\equiv q_y^{eq,*}=-j_y, \widehat{f}_7^{eq,*} \equiv p_{xx}^{eq,*}=
\frac{(j_x^2-j_y^2)}{\gamma\rho}, \widehat{f}_8^{eq,*} \equiv
p_{xy}^{eq,*}=\frac{j_xj_y}{\gamma\rho}$. The definition of the
components of the equilibrium moments are provided in
Appendix~\ref{app:momentcomponents}.

For the D3Q19 model, they become: $ \widehat{f}_0^{eq,*} = \rho,
\widehat{f}_1^{eq,*} \equiv
e^{eq,*}=-11\rho+19\frac{\overrightarrow{j}\cdot\overrightarrow{j}}{\gamma\rho},\newline
\widehat{f}_2^{eq,*} \equiv
e^{2,eq,*}=3\rho-\frac{11}{2}\frac{\overrightarrow{j}\cdot\overrightarrow{j}}{\gamma\rho},
\widehat{f}_3^{eq,*} = j_x,\widehat{f}_4^{eq,*} \equiv
q_x^{eq,*}=-\frac{2}{3}j_x,\widehat{f}_5^{eq,*} = j_y,
\widehat{f}_6^{eq,*} \equiv q_y^{eq,*}=-\frac{2}{3}j_y,
\widehat{f}_7^{eq,*} = j_z, \widehat{f}_8^{eq,*} \equiv
q_z^{eq,*}=-\frac{2}{3}j_z, \widehat{f}_9^{eq,*} \equiv
3p_{xx}^{eq,*}=\frac{\left[3j_x^2-\overrightarrow{j}\cdot\overrightarrow{j}
\right]}{\gamma\rho},\widehat{f}_{10}^{eq,*} \equiv
3\pi_{xx}^{eq,*}=3\left(-\frac{1}{2}p_{xx}^{eq,*}
\right),\widehat{f}_{11}^{eq,*} \equiv
p_{ww}^{eq,*}=\frac{\left[j_y^2-j_z^2
\right]}{\gamma\rho},\widehat{f}_{12}^{eq,*} \equiv
\pi_{ww}^{eq,*}=-\frac{1}{2}p_{ww}^{eq,*},\widehat{f}_{13}^{eq,*}
\equiv
p_{xy}^{eq,*}=\frac{j_xj_y}{\gamma\rho},\widehat{f}_{14}^{eq,*}
\equiv
p_{yz}^{eq,*}=\frac{j_yj_z}{\gamma\rho},\widehat{f}_{15}^{eq,*}
\equiv
p_{xz}^{eq,*}=\frac{j_xj_z}{\gamma\rho},\widehat{f}_{16}^{eq,*} =
0,\widehat{f}_{17}^{eq,*} = 0,\widehat{f}_{18}^{eq,*} = 0$.

A general observation is that only the non-linear terms in the
components of the equilibrium moments are preconditioned by the
parameter $\gamma$. This is consistent with the argument that the
hydrodynamic convective effects, which are non-linear, emerge from
relaxation process during collisions should be contained in these
terms; they should be preconditioned to match the faster propagation
of density perturbations, which are represented by linear terms in
the equilibrium moments. It may be noted that an alternative approach
to preconditioning the equilibria has been proposed recently~\cite{izquierdo08}.

The preconditioned components of the source terms
\begin{equation}
\mathbf{\widehat{S}}^*=\left[
\widehat{S}_0^*,\widehat{S}_1^*,\widehat{S}_2^*,\ldots,\widehat{S}_{b}^*
\right]^{\dag}
\end{equation}
can be written, for the D2Q9 model as: $\widehat{S}_0^* = 0,
\widehat{S}_1^* = 6\frac{(F_xu_x+F_yu_y)}{\gamma^2}, \widehat{S}_2^*
= -6\frac{(F_xu_x+F_yu_y)}{\gamma^2}, \widehat{S}_3^*= \frac{F_x}{\gamma},
\widehat{S}_4^* =-\frac{F_x}{\gamma}, \widehat{S}_5^*=\frac{F_y}{\gamma},
\widehat{S}_6^* =-\frac{F_y}{\gamma}, \widehat{S}_7^*=2\frac{(F_xu_x-F_yu_y)}{\gamma^2},
\widehat{S}_8^*=\frac{(F_xu_y+F_yu_x)}{\gamma^2}$.

The corresponding components of $\mathbf{\widehat{S}}^*$ for the
D3Q19 model are: $\widehat{S}_0^* = 0, \widehat{S}_1^* =
38\frac{(F_xu_x+F_yu_y+F_zu_z)}{\gamma^2}, \widehat{S}_2^* =
-11\frac{(F_xu_x+F_yu_y+F_zu_z)}{\gamma^2}, \widehat{S}_3^*= \frac{F_x}{\gamma},
\widehat{S}_4^* =-\frac{2}{3}\frac{F_x}{\gamma}, \widehat{S}_5^*=\frac{F_y}{\gamma},
\widehat{S}_6^* =-\frac{2}{3}\frac{F_y}{\gamma}, \widehat{S}_7^*=\frac{F_z}{\gamma},
\widehat{S}_8^* = -\frac{2}{3}\frac{F_z}{\gamma}, \widehat{S}_9^* =
2\frac{(2F_xu_x-F_yu_y-F_zu_z)}{\gamma^2},
\newline \widehat{S}_{10}^* =
-\frac{(2F_xu_x-F_yu_y-F_zu_z)}{\gamma^2}, \widehat{S}_{11}^* =
2\frac{(F_yu_y-F_zu_z)}{\gamma^2}, \newline \widehat{S}_{12}^* =
-\frac{(F_yu_y-F_zu_z)}{\gamma^2}, \widehat{S}_{13}^* =
\frac{(F_xu_y+F_yu_x)}{\gamma^2},
\widehat{S}_{14}^* = \frac{(F_yu_z+F_zu_y)}{\gamma^2},
\widehat{S}_{15}^* = \frac{(F_xu_z+F_zu_x)}{\gamma^2}, \widehat{S}_{16}^* = 0,
\widehat{S}_{17}^* = 0, \widehat{S}_{18}^* = 0$.

The preconditioning of the moment projections of the source terms
may also be compactly written as
\begin{equation}
\mathbf{\widehat{S}}^*=\mathcal{P_S}\mathbf{\widehat{S}},
\label{eq:precondsourceterm}
\end{equation}
where
\begin{equation}
\mathcal{P_S}=\diag\left(1,\frac{1}{\gamma^2},\frac{1}{\gamma^2},
\frac{1}{\gamma},\frac{1}{\gamma},\frac{1}{\gamma},\frac{1}{\gamma},\frac{1}{\gamma^2},\frac{1}{\gamma^2}\right)\nonumber
\end{equation}
for the D2Q9 model, and
\begin{equation}
\mathcal{P_S}=\diag\left(1,\frac{1}{\gamma^2},\frac{1}{\gamma^2},
\frac{1}{\gamma},\frac{1}{\gamma},\frac{1}{\gamma},\frac{1}{\gamma},\frac{1}{\gamma},\frac{1}{\gamma},
\frac{1}{\gamma^2},\frac{1}{\gamma^2},\frac{1}{\gamma^2},\frac{1}{\gamma^2},\frac{1}{\gamma^2},\frac{1}{\gamma^2},\frac{1}{\gamma^2},1,1,1\right)\nonumber
\end{equation}
for the D3Q19 model, where the components of the unpreconditioned
source terms $\mathbf{\widehat{S}}$ are given in
Appendix~\ref{app:momentcomponents}.

Clearly, the external forces have a first-order effect on the
convective motion of the fluid, and thus to ``condition" the moments
linearly influenced by such forces, the moment projections need to
be preconditioned by the inverse of $\gamma$. On the other hand,
other moments are effected by the external forces at second order.
These include the ``work" contribution by their interaction with the
fluid motion on the moment corresponding to kinetic energy (see
Appendix~\ref{app:momentcomponents}). Similarly, the moment
projections of the source terms for the momentum flux tensors have
second-order influence. In general, the Chapman-Enskog analysis
reveals that all higher order moments that involve non-linear
effects from interaction of external forces and fluid motion are
much slower than the fluid motion itself and needs to be preconditioned
by the inverse of the square of the preconditioning parameter, i.e.
$1/\gamma^2$ (see Appendix~\ref{app:Chapman_Enskog_pGLBE}).

For the preconditioned collision relaxation time matrix,
\begin{equation}
\widehat{\Lambda}^{*}=\diag(s_0^*,s_1^*,\ldots,s_{b}^*),
\end{equation}
some of the relaxation times $s_{\alpha}^{*}$, i.e. those
corresponding to hydrodynamic modes, can be related to the transport
coefficients. The rest, i.e. those for the kinetic modes, can be
chosen through a von Neumann stability analysis of the linearized
GLBE~\cite{lallemand00,dhumieres02}. For the D2Q9 model, we have
$\frac{1}{s_{1}^*}=3\frac{\zeta}{\gamma}+\frac{1}{2}$, where $\zeta$
is the bulk viscosity, and $s_{7}^*=s_{8}^*=s_{\nu}^*$, where $
\frac{1}{s_{\nu}^*}= 3\frac{\nu}{\gamma}+\frac{1}{2}$, with $\nu$
being shear viscosity. For the kinetic modes, we
have~\cite{lallemand00} $s_1^*=1.63$, $s_2^*=1.14$ and
$s_4^*=s_6^*=1.92$. On the other hand, for the D3Q19
model~\cite{dhumieres02}, we have, for the hydrodynamic modes,
$\frac{1}{s_{1}^*}=\frac{9}{2}\frac{\zeta}{\gamma}+\frac{1}{2}$,
$s_{9}^*=s_{11}^*=s_{13}^*=s_{14}^*=s_{15}^*=s_{\nu}^*$, where $
\frac{1}{s_{\nu}^*}= 3\frac{\nu}{\gamma}+\frac{1}{2}$ and for the
kinetic modes~\cite{dhumieres02}, $s_1^*=1.19$,
$s_2^*=s_{10}^*=s_{12}^*=1.4$, $s_4^*=s_6^*=s_8^*=1.2$ and
$s_{16}^*=s_{17}^*=s_{18}^*=1.98$.

The hydrodynamic fields, i.e., the density $\rho$, velocity
$\overrightarrow{u}$ and pressure $p$ obtained from the solution of
preconditioned GLBE, satisfy the equivalent preconditioned compressible NSE
(see Appendix~\ref{app:Chapman_Enskog_pGLBE}), and can be written as
\begin{equation}
\rho =\sum_{\alpha=0}^{b} f_{\alpha}, \quad \overrightarrow{j}\equiv
\rho\overrightarrow{u} =\sum_{\alpha=0}^{b}
f_{\alpha}\overrightarrow{e}_{\alpha}+\frac{1}{2}\frac{\overrightarrow{F}}{\gamma}\delta_t,
\quad p=c_s^{*2}\rho, \label{eq:precondhydrofields}
\end{equation}
where, $c_s^*=\sqrt{\gamma}c_s$ with $c_s=c/\sqrt{3}$. The
preconditioning of the GLBE effectively reduces the speed of sound
by a factor $\sqrt{\gamma}$. As a result, the disparity between the
propagation speed of density perturbation and that of fluid motion
is decreased by decreasing the parameter $\gamma$. Moreover, the
``effective" Mach number after preconditioning is
$\Ma^*=u/c_s^*=\Ma/\sqrt(\gamma)$. It may be noted that a
Chapman-Enskog analysis of the GLBE carried out in
Appendix~\ref{app:Chapman_Enskog_pGLBE}) also shows how the
evolution of kinetic modes, in addition to the hydrodynamic modes of
interest, are affected by preconditioning. It is evident that the
structure of the preconditioned GLBE with forcing term is very
similar to that without preconditioning, involving only local
scaling of the equilibrium moments, the moment projections of source
terms and the relaxation matrix. As a result, the optimized
computational procedure for GLBE with forcing term described in the
previous section can be fully exploited for the preconditioned
version.

\subsection{\label{sec:srtlbe}Limiting Form: Preconditioned SRT-LBE with Forcing Term}
When all the relaxation parameters are set to the same constant,
i.e.~$s_{\alpha}=1/\tau^*$, we arrive at the SRT-LBE, which can be
conveniently written as the following collision and streaming steps,
respectively, where both steps are expressed in particle velocity
space:
\begin{equation}
\widetilde{f}_{\alpha}(\overrightarrow{x},t)=
-\frac{1}{\tau^*}\left( f_{\alpha}-f_{\alpha}^{eq,*}
\right)+\left(1-\frac{1}{2\tau^*}
\right)S_{\alpha}^* \delta_t,
\end{equation}
where $f_{\alpha}\equiv f_{\alpha}(\overrightarrow{x},t)$, $f_{\alpha}^{eq,*} \equiv f_{\alpha}^{eq,*}(\overrightarrow{x},t)$, and $S_{\alpha}^* \equiv S_{\alpha}^*(\overrightarrow{x},t)$, and
\begin{equation}
f_{\alpha}(\overrightarrow{x}+\overrightarrow{e_{\alpha}}\delta_t,t+\delta_t)=\widetilde{f}_{\alpha}(\overrightarrow{x},t).
\end{equation}
Here, $f_{\alpha}^{eq,*}$ is the preconditioned equilibrium
distribution
\begin{equation}
f_{\alpha}^{eq,*}=w_{\alpha}\left[1+3\overrightarrow{e_{\alpha}}\cdot\overrightarrow{u}+
\frac{9}{2\gamma}(\overrightarrow{e_{\alpha}}\cdot\overrightarrow{u})^2-\frac{1}{2\gamma}\overrightarrow{u}\cdot \overrightarrow{u}\right].
\end{equation}
The hydrodynamic fields can be obtained from the distribution
functions in the same manner as before, i.e. from
Eq.~(\ref{eq:precondhydrofields}). One important consideration is in
obtaining the correct expression for the corresponding
preconditioned source terms. They can be obtained simply by an
inverse transformation of the moment projections of preconditioned
source terms from Eq.~(\ref{eq:precondsourceterm}). That is,
\begin{equation}
\mathbf{S}^*=\mathcal{T}^{-1}\mathbf{\widehat{S}}^*=\mathcal{T}^{-1}\mathcal{P_S}\mathbf{\widehat{S}}.
\label{eq:precondsourcetermvelspace}
\end{equation}
Explicit evaluation of this equation,
Eq.~(\ref{eq:precondsourcetermvelspace}), yields
\begin{equation}
S_{\alpha}^* = w_{\alpha}\left[
3\frac{\left(\overrightarrow{e}_{\alpha}-\frac{\overrightarrow{u}}{\gamma}\right)\cdot\overrightarrow{F}}{\gamma}+
            9\frac{\left(  \overrightarrow{e_{\alpha}} \cdot  \overrightarrow{u} \right)\overrightarrow{e}_{\alpha
            }\cdot\overrightarrow{F}}{\gamma^2}\right],
            \label{eq:precondsourcecorrect}
\end{equation}
which is the desired expression for the source term of the SRT-LBE with
preconditioning. It may be noted that it is essential to maintain the
above form of preconditioned source term to correctly recover the
corresponding preconditioned hydrodynamic behavior. On the other
hand, for example, if, one na\"{i}vely sets
\begin{equation}
S_{\alpha}^* = w_{\alpha}\frac{\left[
3\left(\overrightarrow{e}_{\alpha}-\overrightarrow{u}\right)\cdot\overrightarrow{F}+
            9\left(  \overrightarrow{e_{\alpha}} \cdot  \overrightarrow{u} \right)\overrightarrow{e}_{\alpha
            }\cdot\overrightarrow{F}\right]}{\gamma},
            \label{eq:precondsourceincorrect}
\end{equation}
a Chapman-Enskog analysis of the resulting SRT-LBE (not explicitly
shown here, for brevity) yields macrodynamical equations with
non-vanishing spurious terms (with $\gamma<1$). The $i$-th Cartesian
component of these extra spurious terms to the corresponding
preconditioned momentum equations turns out to be
\begin{equation}
\mbox{Extra
Term}_i=\partial_j\left[\frac{\left(\gamma-1\right)}{\gamma^2}
\left(\tau^*-\frac{1}{2}\right)\delta_t\left(F_iu_j+F_ju_i\right)
\right]
\end{equation}
These terms can indeed dominate with particularly strong
preconditioning at lower $\gamma$, when $(\gamma-1)/\gamma^2$ can
become very large, particularly for spatially and temporally
dependent external forces. For example, simulation of MHD problems,
where Lorentz forces can vary both in space and time, using a
preconditioned SRT-LBE with Eq.~(\ref{eq:precondsourcecorrect})
yielded accurate results, but with
Eq.~(\ref{eq:precondsourceincorrect}), it resulted in grossly wrong
behavior. This stresses the critical importance of properly
preconditioning forcing terms, as the temporal change in the effect
of the external forces on various physical processes during
collisional relaxation are different. In this regard, analysis of
their contributions in moment space, as shown above, is particularly
revealing: the individual contributions of the external forces
spanned in the moment space need to be separately preconditioned
depending on the nature of their effects on the moments.

\section{\label{sec:pvlbe}Preconditioned Vector Lattice Kinetic Equation for Magnetic Induction}
As an illustration of preconditioning an extended system of LBE for
complex fluid flows subjected to external forces, we will now
discuss preconditioning lattice kinetic equations for the magnetic
induction equation required for simulation of MHD flows. Dellar~\cite{dellar02}
concluded that a vector formulation of the kinetic equation is necessary to properly
recover the magnetic induction equation and constructed a 2D model to accomplish this,
which was extended to 3D by Breyannis and Valeougeorgis~\cite{breyiannis04}.
The GLBE with forcing term is used in conjunction with such a lattice kinetic
equation, with the latter providing the Lorentz force field to the former.

In addition to the propagation of the density perturbation as sound
with speed $c_s$, MHD flows are characterized by the propagation of
perturbation of magnetic induction, the so-called Alfv\'{e}n waves. If
$B_i$ is the Cartesian component of magnetic induction, we can
obtain the corresponding Alfv\'{e}n velocity as $V_{A,i}=B_i/\sqrt{\rho
\mu}$, where $\rho$ and $\mu$ are the density and magnetic
permeability, respectively. Thus, we can define a local Alfv\'{e}n
number $\Al=V_{A,i}/c_s$, and Dellar~\cite{dellar02}
constructed a lattice kinetic equation that recovers the magnetic
induction equation applicable at low $\Al$, with deviations
$O(\Al^2)$. In this scaling, $O(V_{A,i})\approx O(u_i)$, or
$O(\Al)\approx O(\Ma)$. Thus, in MHD flows, there is an additional
disparity between the speed of perturbation of the magnetic induction
field and the the speed of sound, The condition number, in this
case, is inversely proportional to the Alfv\'{e}n number, i.e.
$O(1/\Al)$. Preconditioning the lattice kinetic equation accelerates
its steady state convergence by reducing the disparity between such
characteristic speeds in MHD flows.

We now develop a preconditioning formulation for a unified vector lattice
kinetic equation for magnetic induction applicable in both 2D and
3D. Unlike the case of fluid flow, which has fourth-order isotropy
requirements on lattice velocity models to correctly recover viscous
stress tensor, the magnetic induction imposes lower order symmetry
requirements. Thus, we need only a smaller number of particle
velocity directions for magnetic induction
$\overrightarrow{e_{\alpha}}^m$, and following previous work,
we consider a D2Q5 model
\begin{equation}
\overrightarrow{e}_{\alpha}^m = \left\{\begin{array}{ll}
   {(0,0)}&{\alpha=0}\\
   {(\pm 1,0),(0,\pm 1)}&{\alpha=1,\cdots,4}\\
\end{array} \right.
\label{eq:velocityd2q5}
\end{equation}
and a D3Q7 model
\begin{equation}
\overrightarrow{e} _{\alpha}^m= \left\{\begin{array}{ll}
   {(0,0,0)}&{ \alpha=0}\\
   {(\pm 1,0,0),(0,\pm 1,0),(0,0,\pm 1)}&{ \alpha=1,\cdots,6}\\
\end{array} \right.
\label{eq:velocityd3q19}
\end{equation}
in 2D and 3D, respectively. The preconditioned lattice kinetic
equation can, then, be written in terms of the following collision
and streaming steps, respectively:
\begin{equation}
\widetilde{g}_{\alpha j}(\overrightarrow{x},t)=
-\frac{1}{\tau_m^*}\left( g_{\alpha j}-g_{\alpha j}^{eq,*}
\right)\label{eq:mcollision}
\end{equation}
where $g_{\alpha j}\equiv g_{\alpha j}(\overrightarrow{x},t)$ and
$g_{\alpha j}^{eq,*}\equiv g_{\alpha j}^{eq,*}(\overrightarrow{x},t)$, and
\begin{equation}
g_{\alpha j}(\overrightarrow{x}+\overrightarrow{e}_{\alpha
j}^m\delta_t,t+\delta_t)=\widetilde{g}_{\alpha
j}(\overrightarrow{x},t),\label{eq:mstreaming}
\end{equation}
where $g_{\alpha j}$ is the vector distribution function in index
notation  $\alpha=0,1,\cdots,b_m$. Here, $b_m=4$ and $b_m=6$ in
2D and 3D, respectively. The subscript Roman indices $i$, $j$, etc.,
represent Cartesian components of the coordinate directions.
Assuming the usual summation convention of repeated indices, the
Cartesian component of the preconditioned vector equilibrium
distribution function $g_{\alpha j}^{eq,*}$ is given as
\begin{equation}
g_{\alpha j}^{eq,*}=W_{\alpha}\left\{B_j+\frac{e_{\alpha
k}^m}{\theta_m}\left(\frac{u_kB_j-B_ku_j}{\gamma_m}\right)\right\},
\label{eq:mageqmoment}
\end{equation}
where
\begin{equation}
W_{\alpha}= \left\{\begin{array}{ll}
   {1/N_m}&{\alpha=0}\\
   {1/(2N_m)}&{\alpha=1,\cdots,b_m}\\
\end{array} \right.
\end{equation}
and $\theta_m=1/N_m$, with $N_m=3$ and $N_m=4$ in 2D and 3D,
respectively. Here, $B_j$ is the Cartesian component of the magnetic
induction and $\gamma_m$ is the preconditioning parameter, with
$0<\gamma_m\leq 1$. Thus, the preconditioning is carried out on the
non-linear part of the vector equilibrium distribution,
Eq.~(\ref{eq:mageqmoment}), which represents the transport of
magnetic induction field by the fluid motion. The preconditioned
relaxation time $\tau_m^*$ is related to the magnetic diffusivity of
the fluid $\eta_m$, where $\eta_m=1/(\mu\sigma)$ with $\mu$ and
$\sigma$ being the magnetic permeability and electrical
conductivity, respectively, and is given as
\begin{equation}
\tau_m^*=\frac{\eta_m}{\gamma_m\theta_m}+\frac{1}{2}.
\label{eq:mrelaxationtime}
\end{equation}

Once the vector distribution function is calculated, the components
of the magnetic induction $B_i$ and the current density $J_i$ can be
obtained by taking their zeroth and first moments as:
\begin{equation}
B_i=\sum_{\alpha=0}^{b_m}g_{\alpha i} \label{eq:magneticinduction}
\end{equation}
and
\begin{equation}
J_i\equiv\frac{1}{\mu}\left(\overrightarrow{\nabla}\times\overrightarrow{B}\right)_i=-\frac{1}{\mu}\frac{1}{\tau_m^*\theta_m}\epsilon_{ijk}\sum_{\alpha=0}^{b_m}\left(e_{\alpha
k} g_{\alpha j}-e_{\alpha k} g_{\alpha j}^{eq,*}\right),
\end{equation}
where $\epsilon_{ijk}$ is the Levi-Civita or the third-order
permutation tensor. It can be shown that the preconditioned vector
lattice kinetic equation can recover the corresponding
preconditioned lattice induction equation given as follows (see
Appendix~\ref{app:Chapman_Enskog_vectorLBE}):
\begin{equation}
\partial_t
B_i+\frac{1}{\gamma_m}\nabla_j\left(u_jB_i-B_ju_i\right)=\frac{1}{\gamma_m}\partial_j\left(
\eta_m
\partial_j B_i \right).
\label{eq:magneticinduction}
\end{equation}
As shown by Dellar~\cite{dellar02}, the magnetic induction
will remain solenoidal, i.e. $\partial_iB_i=0$, provided the initial
condition on the magnetic induction satisfies the divergence free
condition.

The interaction of the magnetic induction and the current density
gives rise to the Lorentz force on the fluid flow. This force can be
written as
\begin{equation}
\overrightarrow{F}_{Lorentz}=\overrightarrow{J}\times\overrightarrow{B}
\end{equation}
and enters as $\overrightarrow{F}=\overrightarrow{F}_{Lorentz}$ in
the preconditioned GLBE discussed in the previous section.

\subsection{\label{subsec:lowmagneticRe}Achieving Low Magnetic
Reynolds Number or Magnetic Prandtl Number at Steady State} If $L$,
$u_0$, $B_0$ are the characteristic length, velocity and magnetic
induction scales, respectively, then we can non-dimensionalize
various quantities as $\overrightarrow{\nabla} \leftarrow
L\overrightarrow{\nabla}$, $t \leftarrow (u_0/L)t$, $u \leftarrow
u/u_0$ and $B \leftarrow B/B_0$, and obtain the non-dimensional form
of magnetic induction equation, Eq.~(\ref{eq:magneticinduction}) as
\begin{equation}
\gamma_m\partial_t
B_i+\nabla_j\left(u_jB_i-B_ju_i\right)=\partial_j\left(
\frac{1}{\Reyn_m}
\partial_j B_i \right),
\label{eq:magneticinduction}
\end{equation}
where $\Reyn_m=u_0L/\eta_m$ is the magnetic Reynolds number and the
corresponding dimensionless current density is
$\overrightarrow{J}=(1/\Reyn_m)\overrightarrow{\nabla}\times\overrightarrow{B}$.

Now, in many practical MHD applications, particularly for liquid
metals, $\Reyn_m$ is relatively very small, or so is the magnetic
Prandtl number $\Pran_m$, which is given by $\Pran_m=\nu / \eta=\Reyn_m / \Reyn$:
$\Reyn_m \ll O(1)$ and $\Pran_m \ll O(1)$. To achieve lower $\Reyn_m$ or
$\Pran_m$, we need to make $\eta_m$ smaller, which, in turn, from
Eq.~(\ref{eq:mrelaxationtime}), means reducing $\tau_m$.
However,  true for a typical lattice based method, as $\tau_m$
approaches $0.5$, it can cause numerical instability. This situation
can be remedied for the steady state situation by considering the
following: at steady state, we have
$\nabla_j\left(u_jB_i-B_ju_i\right)=\partial_j\left( \frac{1}{\Reyn_m}
\partial_j B_i \right)$, to which we apply a scaling factor $\chi$
as $\chi\nabla_j\left(u_jB_i-B_ju_i\right)=\partial_j\left(
\frac{1}{\Reyn_m}
\partial_j B_i \right)$. This effectively changes $\Reyn_m$ to
$\chi \Reyn_m$ at steady state. Thus, we can write $\Reyn_{m,\eff}=\chi
\Reyn_m$, resulting in $\Pran_{m,\eff}=\chi \Pran_m$. In the context of
preconditioning, a lattice kinetic scheme for this ``effective"
steady state magnetic induction equation can readily be constructed.

Thus, in dimensional form, we need to construct a scaled lattice
kinetic scheme with preconditioning for the following macroscopic
equation
\begin{equation}
\partial_t
B_i+\frac{\chi}{\gamma_m}\nabla_j\left(u_jB_i-B_ju_i\right)=\frac{1}{\gamma_m}\partial_j\left(
\eta_m\partial_j B_i \right). \label{eq:effmequation}
\end{equation}
The magnetic field satisfying this equation,
Eq.~(\ref{eq:effmequation}), can be obtained by solving the above
preconditioned lattice kinetic scheme, i.e.
Eqs.~(\ref{eq:mcollision}) and (\ref{eq:mstreaming}), with its
associated auxiliary equations, except for the following changes in
the computation of vector equilibrium distribution and current
density:
\begin{equation}
g_{\alpha j}^{eq,*}=W_{\alpha}\left\{B_j+\frac{e_{\alpha
k}^m}{\theta_m}\frac{\chi}{\gamma_m}\left(u_kB_j-B_ku_j\right)\right\}
\end{equation} and
\begin{equation}
J_i\equiv\frac{1}{\mu\chi}\left(\overrightarrow{\nabla}\times\overrightarrow{B}\right)_i=
-\frac{1}{\mu\chi}\frac{1}{\tau_m^*\theta_m}\epsilon_{ijk}\sum_{\alpha=0}^{b_m}\left(e_{\alpha
k} g_{\alpha j}-e_{\alpha k} g_{\alpha j}^{eq,*}\right).
\end{equation}

\section{\label{sec:results}Results and Discussion}
We will now present investigations of the preconditioned
computational approach presented in the previous sections by means
of a set of canonical examples. Unless otherwise stated, all the
results will be expressed in the natural lattice units of the
method, i.e. we use the lattice spacing $\delta_x$ as the length
scale and the particle velocity $c$ as the velocity scale (with
$\delta_x/c$ used to scale the temporal quantities).

\begin{figure}
\includegraphics[width = 105mm]{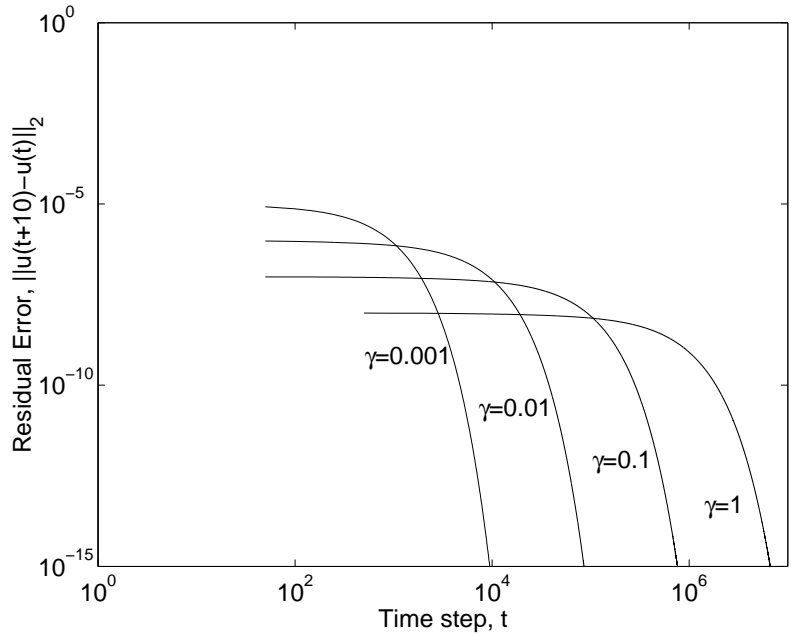}
\caption{\label{fig:Poiseuille_Reserror_nu001} Convergence histories
(in log-log scale) of preconditioned MRT-LBE with forcing term for
simulation of Poiseuille flow with maximum flow velocity
$u_{max}=0.00051$ and kinematic viscosity $\nu=0.001$; $\Reyn=32.6$.}
\end{figure}

First, we simulate the simple classical problem of flow with a fluid
viscosity $\nu$ between parallel plates spaced $2L$ apart and
driven by a pressure gradient $-dp/dx$, i.e. plane Poiseuille flow
using the preconditioned GLBE with forcing term. We consider the
domain to be periodic in the streamwise and spanwise directions, and
thus the pressure gradient is applied as a body force. No slip
conditions at the walls are specified by using the half-way or link
bounce back scheme~\cite{ladd94}. For this setup, if
$\overrightarrow{F}=-(dp/dx)\widehat{i}$ is the driving force, the
maximum fluid velocity occurring at the center is
$u_{max}=FL^2/(2\rho_0\nu)$, where $\rho_0$ is the nominal fluid
density. We set $L=32$, $\nu=0.001$ and $\rho_0=1$, and apply a
pressure gradient such that the maximum velocity is
$u_{max}=0.00051$, or $\Ma=0.0008333$. The Reynolds number based on
the above velocity and $L$ becomes $\Reyn=32.6$.
Figure~\ref{fig:Poiseuille_Reserror_nu001} shows the convergence
history with different values of preconditioning parameter $\gamma$
(no preconditioning corresponds to $\gamma=1$) for this external
force driven flow problem with relatively low $\Ma$.
The convergence to steady state is measured by the second-norm of
the residual error of the velocity, $||u(t+10)-u(t)||_2$. It can be
seen that preconditioning the GLBE with forcing term dramatically
accelerates the steady state convergence for this problem, in
particular, by more than two orders of magnitude with strong
preconditioning carried out using $\gamma=0.001$.

\begin{figure}
\includegraphics[width = 105mm]{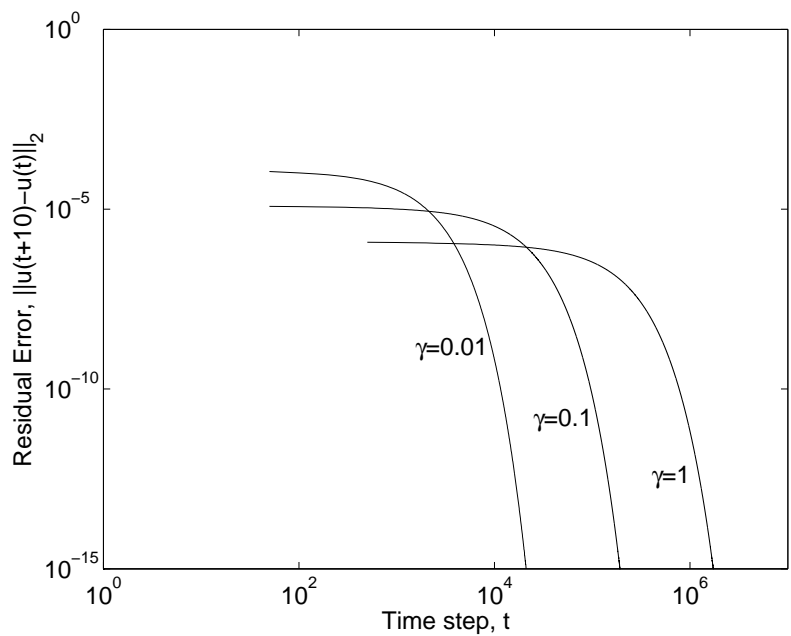}
\caption{\label{fig:Poiseuille_Reserror_nu005} Convergence histories
(in log-log scale) of preconditioned MRT-LBE with forcing term for
simulation of Poiseuille flow with maximum flow velocity
$u_{max}=0.0128$ and kinematic viscosity $\nu=0.005$; $\Reyn=163.8$.}
\end{figure}

\begin{figure}
\includegraphics[width = 105mm]{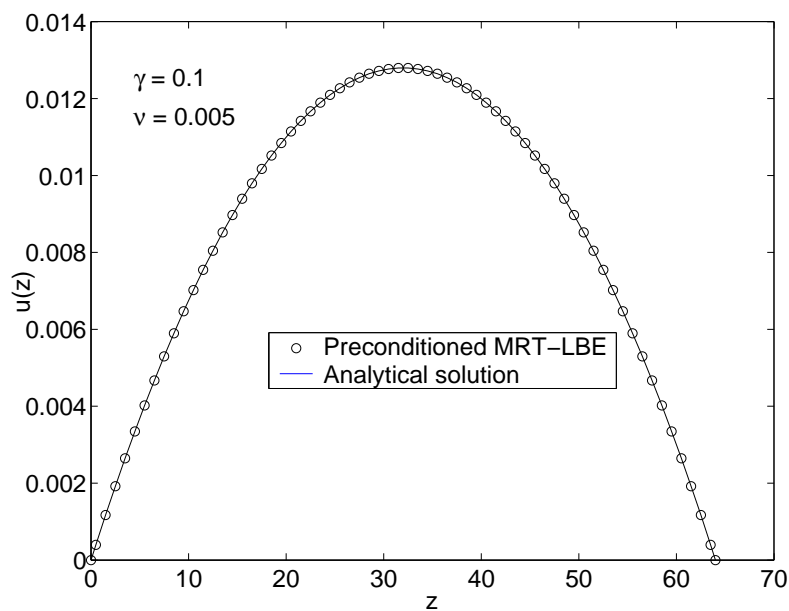}
\caption{\label{fig:Poiseuille_CompareSolution} Comparison of
computed velocity profile using the preconditioned MRT-LBE with
analytical solution for simulation of Poiseuille flow with
preconditioning parameter $\gamma=0.1$ and kinematic viscosity
$\nu=0.005$; $\Reyn=163.8$}
\end{figure}

Next, we simulate plane Poiseuille flow with relatively higher $\Ma$
and $\Reyn$. As before, we set $L=64$, but use $\nu=0.005$ and apply a
pressure gradient such that $\Ma=0.00222$ and $\Reyn=163.8$. The
convergence history for this problem is presented in
Fig.~\ref{fig:Poiseuille_Reserror_nu005}.
Again, a significant reduction in the number of time steps for
convergence to steady state is achieved through preconditioning. On
the other hand, it appears that to maintain numerical stability the
minimum possible value of the preconditioning parameter $\gamma$
needs to be higher at higher $\Ma$. The computed velocity profile for
problem with preconditioning using $\gamma=0.1$ compared with the
analytical solution is shown in
Fig.~\ref{fig:Poiseuille_CompareSolution}. Excellent agreement is
seen.

\begin{figure}
\includegraphics[width = 105mm]{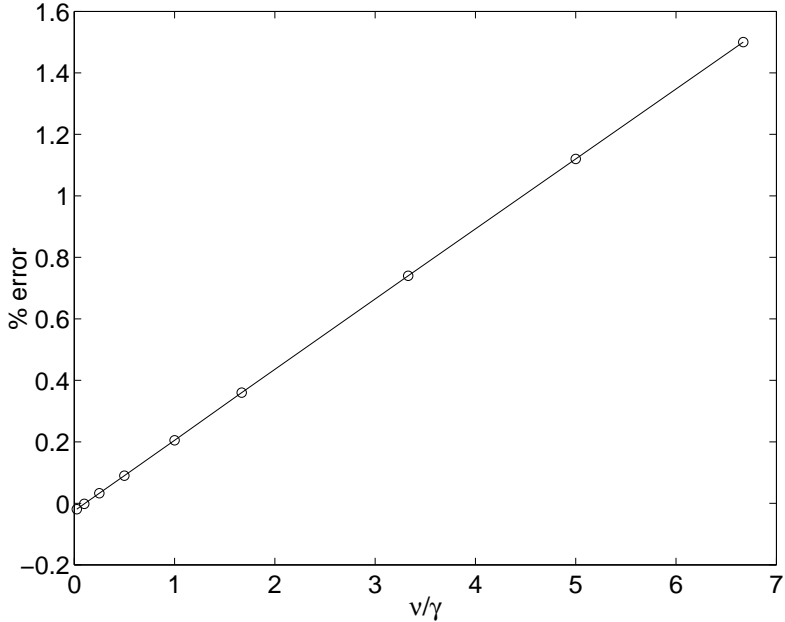}
\caption{\label{fig:Poiseuille_error} Error in the computed velocity
at the center by using the preconditioned MRT-LBE for simulation of
Poiseuille flow as a function of the ratio $\nu/\gamma$.}
\end{figure}

When all the other parameters are maintained constant, the deviation
of the computed solution from the analytical solution is related to
the ratio $\nu/\gamma$, which, in turn, is related to the
hydrodynamic relaxation time in the preconditioned GLBE as
$\frac{1}{s_{\nu}^*}=3\frac{\nu}{\gamma}+\frac{1}{2}$ (see
Sec.~\ref{subsec:pglbe}). Figure~\ref{fig:Poiseuille_error}
shows the relative error in the computed velocity as a function of
$\nu/\gamma$. It is evident that the error, which remains relatively
small, is linearly proportional to this ratio. Thus, by maintaining
low values of $\nu/\gamma$, the error can be correspondingly kept
small.

\begin{figure}
\includegraphics[width = 105mm]{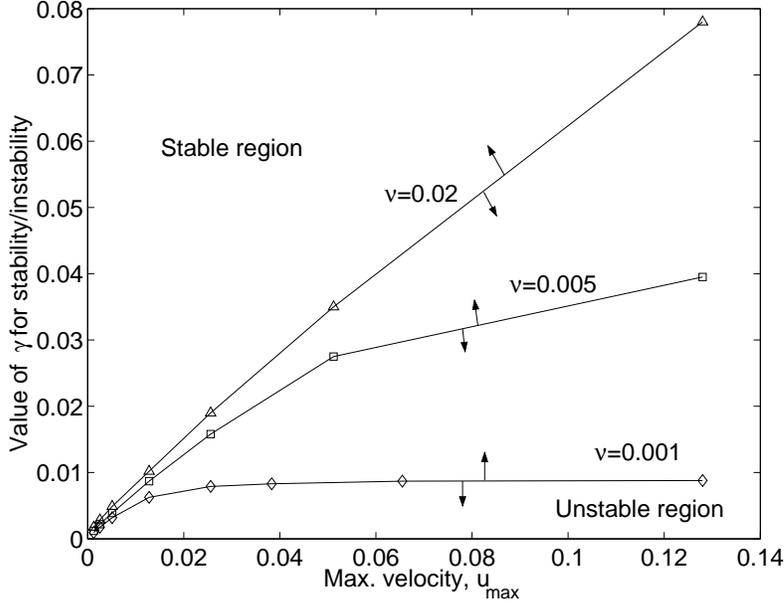}
\caption{\label{fig:Poiseuille_stabilityspace} Parameter space of
the preconditioning parameter $\gamma$ for stable/unstable
computations using the preconditioned MRT-LBE for simulation of
Poiseuille flow at different maximum possible flow velocities
$u_{max}$ and kinematic viscosities $\nu$ of the fluid.}
\end{figure}

We now investigate the influence of preconditioning the GLBE on
numerical stability for the case of plane Poiseuille flow considered
above. This can be done by systematically carrying out simulations
at different characteristic parameters, including $\gamma$, $\nu$
and $\Ma$, and determine the threshold at which the computations
become unstable, i.e. small variations growing exponentially with time.
The results can be conveniently expressed in the form of
a regime map or parameter space that delineates stable and unstable
parameter sets. Figure~\ref{fig:Poiseuille_stabilityspace} shows the
stability-instability parameter space determined by the maximum flow
velocity $u_{max}$ and the preconditioning parameter $\gamma$ for
different fluid viscosity $\nu$.
Arrows normal to the curves pointed upwards indicate stable
parameter space, while those pointed downwards pertain to unstable
space. This regime map is particularly revealing. First, for a given
fluid viscosity $\nu$, as the maximum velocity or, equivalently,
$\Ma$ is reduced, lower values of the preconditioning parameter
$\gamma$ can be used, i.e. the GLBE can be strongly preconditioned
resulting in greater computational gains while maintaining numerical
stability. The fluid viscosity appears to significantly affect the
stability parameter space. For a given $u_{max}$, the minimum
possible value of $\gamma$ is higher at higher values of $\nu$. That
is, the extent of preconditioning is greater with lower fluid
viscosities. Interestingly, Fig.~\ref{fig:Poiseuille_stabilityspace}
also shows that the delineating curve has a linear
functional relationship between $\gamma$ and $u_{max}$ at higher
$\nu$, while at lower $\nu$, the stability envelope is nearly flat with
a constant $\gamma$ for a wide variation of $u_{max}$. Thus, in
general, the benefits of preconditioning, while maintaining
numerical stability, is greater at lower $\Ma$ and lower $\nu$. It
may be noted that the use of GLBE in lieu of SRT-LBE in the context
of preconditioning results in significant enhancement of numerical
stability, as will be shown later for fully 3D problems
characterized by complex fluid motions.

Let us now consider a problem in which we can investigate
preconditioning a system of LBE, where the external force impressed
on the fluid is a also a strong function of the fluid motion itself.
This is, we consider the classical Hartmann flow, consisting of
pressure driven flow between two parallel plates in the presence of
a magnetic field perpendicular to the walls. In addition to the
Reynolds number $\Reyn$, this problem is characterized by the Hartmann
number $\Ha$, which is given by
$\Ha=B_0L\sqrt{\frac{\sigma}{\rho\nu}}$, where $B_0$ and $\sigma$ are
the applied magnetic field strength and fluid's electrical
conductivity, respectively, and other parameters are as previously
defined.

We now simulate this problem by using a vector lattice kinetic
scheme preconditioned by parameter $\gamma_m$ for magnetic induction
(as in Sec.~\ref{sec:pvlbe}), in conjunction with the GLBE with forcing
term preconditioned by parameter $\gamma$ (as in
Sec.~\ref{subsec:pglbe}). The Lorentz force arising from the
interaction of the magnetic induction and velocity field is
introduced into consistently preconditioned forcing terms. To
simulate the flow of liquid metals at low magnetic Reynolds number
$\Reyn_m$ or magnetic Prandtl number $\Pran_m$, we further apply a scaling
factor $\chi$ to the preconditioned lattice kinetic scheme (as in
Sec.~\ref{subsec:lowmagneticRe}).

\begin{figure}
\includegraphics[width = 105mm]{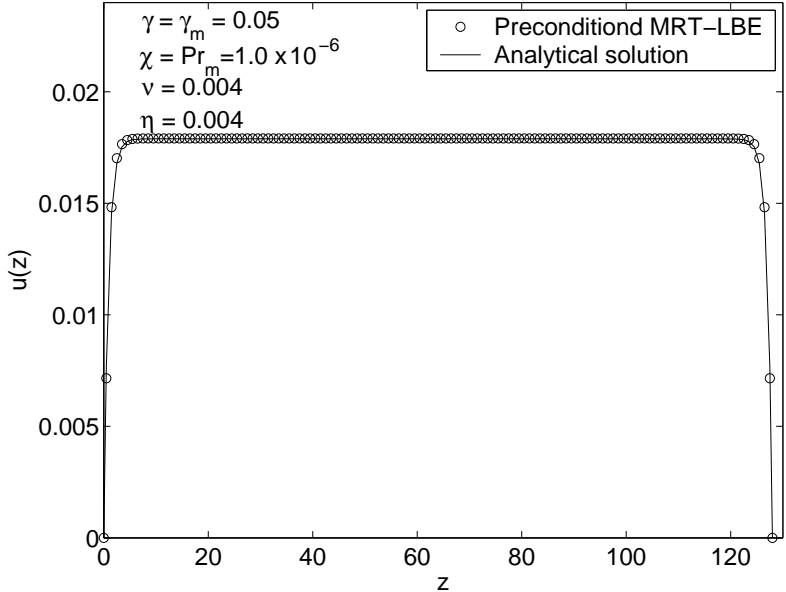}
\caption{\label{fig:Hartmann_CompareV} Comparison of computed
velocity profile using the preconditioned MRT-LBE with analytical
solution for simulation of Hartmann flow with preconditioning
parameters $\gamma=\gamma_m=0.05$ and kinematic viscosity
$\nu=0.004$; $\Reyn=286$, $\Ha=71.6$.}
\end{figure}

\begin{figure}
\includegraphics[width = 105mm]{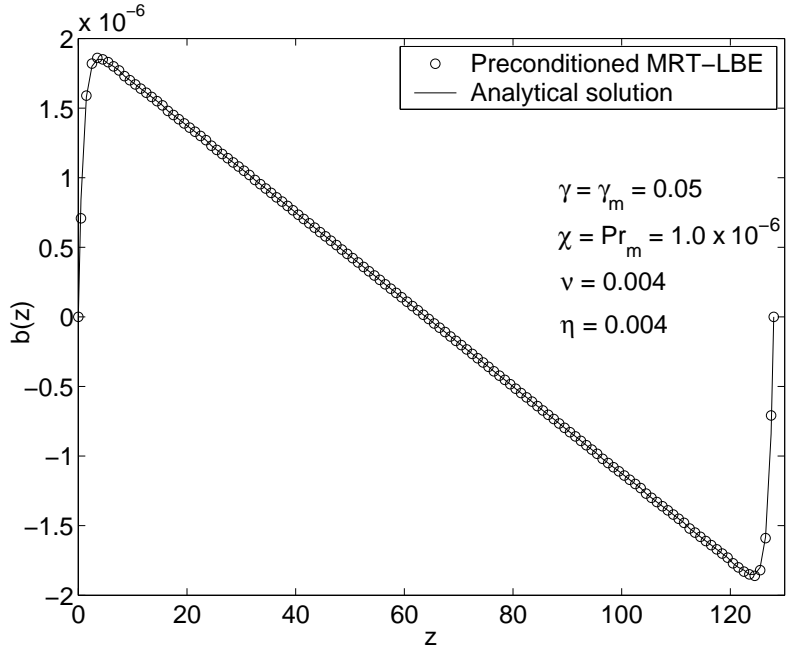}
\caption{\label{fig:Hartmann_CompareB} Comparison of computed
induced magnetic field profile using the preconditioned MRT-LBE with
analytical solution for simulation of Hartmann flow with
preconditioning parameters $\gamma=\gamma_m=0.05$ and kinematic
viscosity $\nu=0.004$; $\Reyn=286$, $\Ha=71.6$.}
\end{figure}

Figure~\ref{fig:Hartmann_CompareV} shows the computed steady state
velocity profile of Hartmann flow with $\Reyn=286$, $\Ha=71.6$,
$\Pran_m=1.0\times 10^{-6}$.
This is achieved by using transport coefficients of $\nu=0.004$, and
$\eta=0.004$, and $L=64$, along with preconditioning and scaling
factors of $\gamma=0.05$, $\gamma_m=0.05$ and
$\chi=1.0~\times~10^{-6}$. The corresponding computed steady state
magnetic induction profile is presented in
Fig.~\ref{fig:Hartmann_CompareB}.
The flattening of the velocity profile observed is characteristic of
MHD flows due to the Lorentz force, with most of the velocity
variation being confined to the region very close to the wall, in
the so-called Hartmann layer. In general, the thickness of Hartmann
layer $\delta_H$ is inversely proportional to the Hartmann number
($\delta_H\sim 1/\Ha$). The analytical solution for this problem is
provided in standard texts on MHD (e.g., Ref.~\cite{muller01}). It
is seen that these profiles computed using preconditioned system of
LBE are in excellent agreement with corresponding analytical
solutions.

\begin{figure}
\includegraphics[width = 105mm]{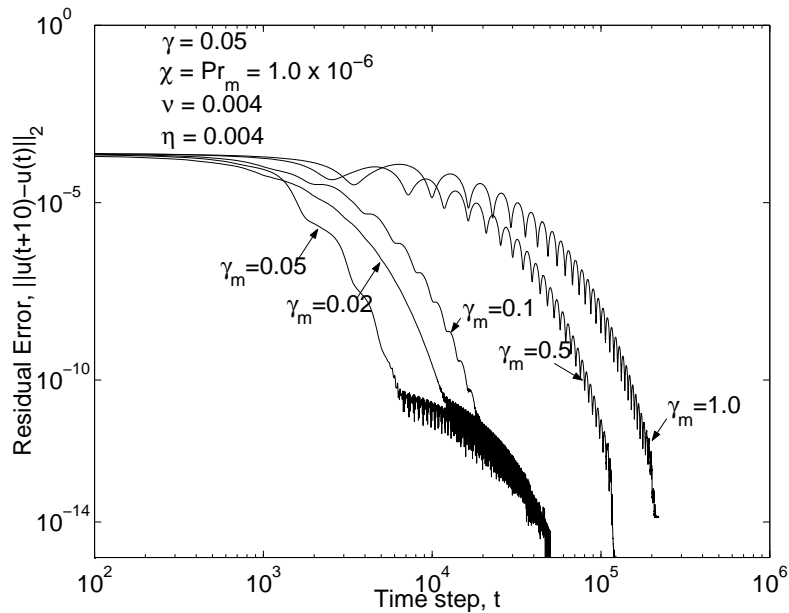}
\caption{\label{fig:Hartmann_Reserror_gamma05} Convergence histories
(in log-log scale) of preconditioned MRT-LBE with forcing term in
conjunction with preconditioned vector lattice kinetic scheme for
simulation of Hartmann flow with flow preconditioning parameter
$\gamma=0.05$ for various values of induction preconditioner
parameter $\gamma_m$; $\Reyn=286$, $\Ha=71.6$.}
\end{figure}

Let us now investigate the behavior of steady state convergence of
this problem by applying various levels of preconditioning.
Figure~\ref{fig:Hartmann_Reserror_gamma05} shows the convergence
history obtained at various values of $\gamma_m$ at a fixed value of
$\gamma$, i.e. $\gamma=0.05$.
Interestingly, preconditioning only the GLBE with forcing term, but
not the vector lattice kinetic scheme (i.e. with $\gamma_m=1.0$)
results in the slowest convergence. However, as we increase the
extent of preconditioning by lowering the values of $\gamma_m$, the
number of time steps to reach steady state is significantly reduced.
The benefits of preconditioning are greatest when both the
preconditioning parameters are equal to one another,
i.e. $\gamma=\gamma_m=0.05$. Moreover, it is also interesting to observe
that if the lattice kinetic scheme is more strongly preconditioned
than that for the GLBE, i.e. $\gamma_m<\gamma$, the approach to steady state
becomes slower as compared with the case with $\gamma_m=\gamma$.
This is consistent with the scaling $O(\Al)\sim
O(\Ma)$, that is, the propagation speeds of both magnetic field and
velocity field by the fluid convective motion occur at similar time
scales. Both these are slower than the speed of density perturbation
and, thus, preconditioning the terms representing these two physical
processes by the same magnitudes would result in the fastest steady
state convergence.

\begin{figure}
\includegraphics[width = 105mm]{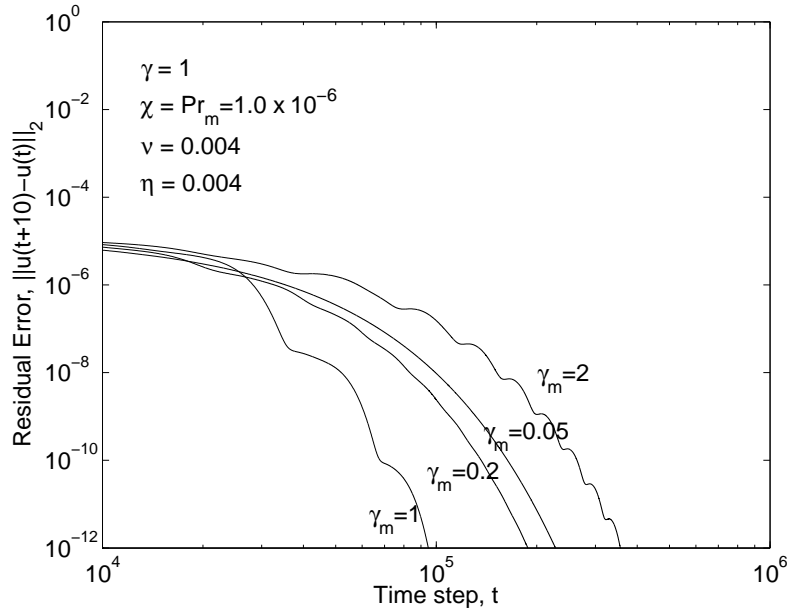}
\caption{\label{fig:Hartmann_Reserror_gamma10} Convergence histories
(in log-log scale) of MRT-LBE with forcing term without
preconditioning in conjunction with preconditioned vector lattice
kinetic scheme for simulation of Hartmann flow with flow
preconditioning parameter $\gamma=1.0$ for various values of
induction preconditioner parameter $\gamma_m$; $\Reyn=286$, $\Ha=71.6$.}
\end{figure}

\begin{figure}
\includegraphics[width = 105mm]{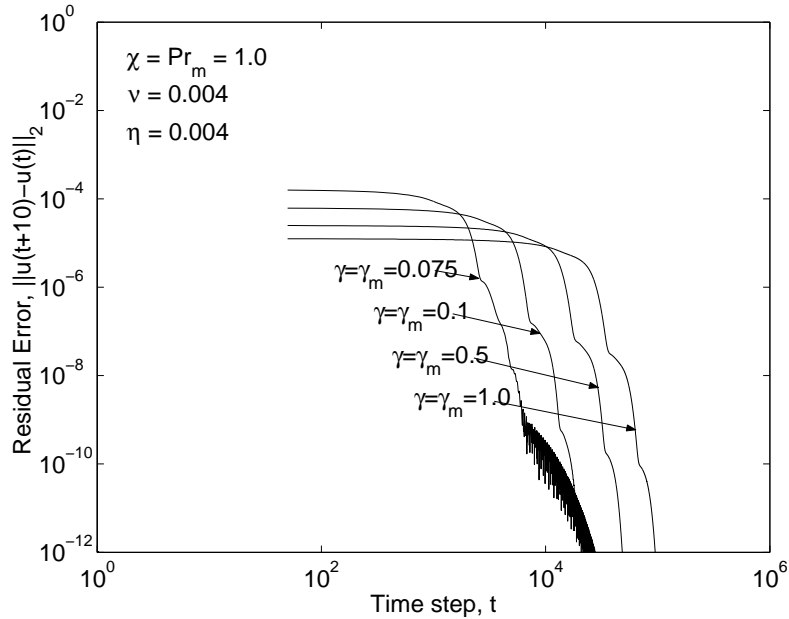}
\caption{\label{fig:Hartmann_Reserror_eqgam_Pr1} Convergence
histories (in log-log scale) of preconditioned MRT-LBE with forcing
term in conjunction with preconditioned vector lattice kinetic
scheme for simulation of Hartmann flow with equal values of flow
($\gamma$) and induction ($\gamma_m$) preconditioning parameters;
$\Reyn=286$, $\Ha=71.6$.}
\end{figure}

This notion is further illuminated by considering cases in which the
GLBE is used without preconditioning ($\gamma=1$), while the lattice
kinetic scheme is solved with different values of the
preconditioning parameter $\gamma_m$. The results with these tests
are plotted in Fig.~\ref{fig:Hartmann_Reserror_gamma10}.
It shows that the convergence rate actually becomes slower if only
one of the LBEs is preconditioned, no matter what the value of the
preconditioning parameter is used, when the other LBE is used
without preconditioning. Thus, preconditioning should be done for
both the LBE and at the same levels. This is further corroborated by
considering a series of cases with $\gamma=\gamma_m$, as shown in
Fig.~\ref{fig:Hartmann_Reserror_eqgam_Pr1}, with lower values for
these parameters providing more rapid convergence to steady state.

It may be noted that, in a recent work, we studied a series of
problems with very thin Hartmann layers by introducing stretched
grids through the Roberts boundary layer
transformation~\cite{roberts71} by means of an
interpolation-supplemented streaming step in the LBE
framework~\cite{pattison07}. In particular, using this modified LBE,
we simulated Hartmann flow with $\Ha$ as high as $10,000$ in very
good comparison with corresponding asymptotic analytical results,
leading to very significant reduction in the number of grid nodes as
compared to using standard LBE using uniform grids~\cite{pattison07}.

\begin{figure}
\includegraphics[width = 100mm,viewport=120 120 580
490,clip]{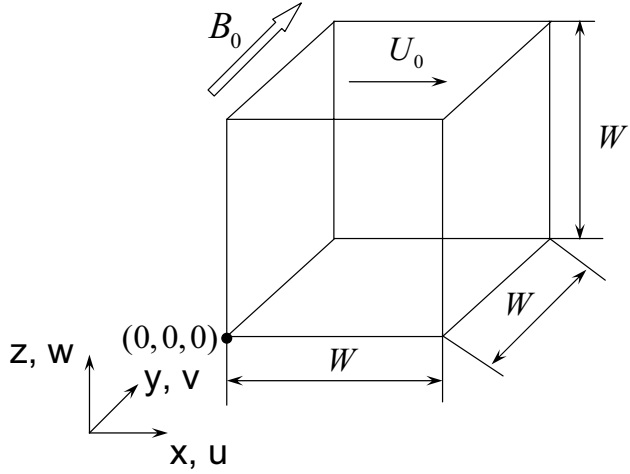}
\caption{\label{fig:3D_Driven_Cavity_Schematic} Schematic
illustration of a three-dimensional (3D) cubical cavity with its top
lid moving at velocity $U_0$ in the presence of an external magnetic
field $B_0$, applied parallel to the lid surface and perpendicular
to its direction of motion.}
\end{figure}

We now present some applications of preconditioned LBE for
simulation of 3D wall bounded shear flows of electrically conducting
fluids such as liquid metals mediated by magnetic fields. In this
regard, we consider a cubic cavity of side length $W$ containing
liquid metal which is driven by its top lid moving at velocity
$U_0$. An external magnetic field $B_0$ is applied parallel to the
lid surface and perpendicular to its direction of motion. A
schematic of this flow problem is shown in
Fig.~\ref{fig:3D_Driven_Cavity_Schematic}.

\begin{figure}
\includegraphics[width = 105mm]{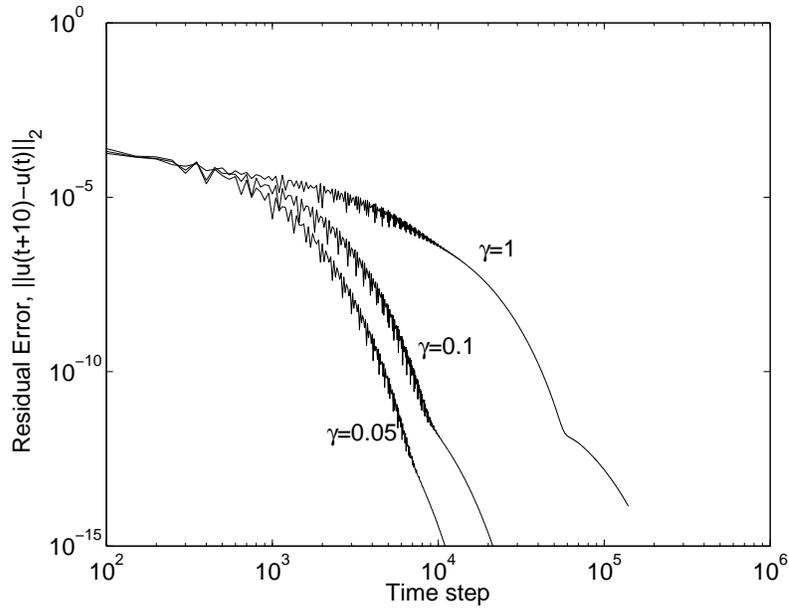}
\caption{\label{fig:3Ddrivencavity_Reserror_noMHD} Convergence
histories (in log-log scale) of preconditioned MRT-LBE with forcing
term for simulation of flow in a 3D cubical driven cavity in the
absence of an applied magnetic field with lid velocity $U_0=0.0235$
and kinematic viscosity $\nu=0.015$ at different values of flow
preconditioning parameters $\gamma$; $\Reyn=100$, grid resolution is
$64^3$.}
\end{figure}

We consider the top lid moving with a velocity $U_0=0.0235$
imparting shear on the fluid of viscosity $\nu=0.015$ contained in a
cavity discretized by $64^3$ grid nodes, such that the Reynolds
number is $100$. Initially, we clarify the advantage of
preconditioning for this highly 3D flow even for a simpler case that
does not involve the application of magnetic field. The velocity
boundary conditions at the walls, including the top moving lid, were
imposed by using a link bounce back scheme~\cite{ladd94}. For the
moving wall, this scheme adds contributions due to appropriate
momentum to the distribution functions. The convergence histories in
the absence of an external magnetic field for different values of
$\gamma$ are shown in Fig.~\ref{fig:3Ddrivencavity_Reserror_noMHD}.
It is seen that significant reduction in the number of time steps to reach
steady sate, is obtained with preconditioning for this problem.

\begin{figure}
\includegraphics[width = 105mm]{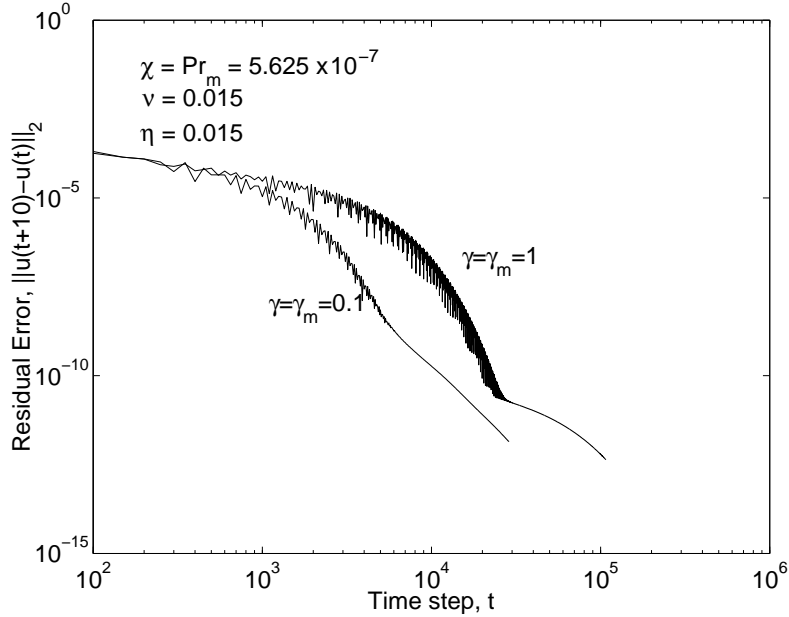}
\caption{\label{fig:3Ddrivencavity_Reserror_MHD} Convergence
histories (in log-log scale) of preconditioned MRT-LBE with forcing
term for simulation of flow in a 3D cubical driven cavity in the
presence of an applied magnetic field with lid velocity $U_0=0.0235$
and kinematic viscosity $\nu=0.015$ at different values of
preconditioning parameters, with $\gamma=\gamma_m$; $\Reyn=100$,
$\Ha=45$, grid resolution is $64^3$.}
\end{figure}

We now consider the case involving the application of an external
magnetic field such that the Hartmann number is $45$, which is
obtained by setting $\eta=0.015$, and the magnetic Prandtl number is
$5.625\times 10^{-7}$. We consider the induced magnetic fields
at far-off distances outside of the cavity to be zero as our
boundary condition. This is achieved by considering a larger
computational domain for the magnetic induction field encompassing
the cavity walls. On these extended computational boundaries, we
implement a zero induced field condition through an extrapolation
method~\cite{chen96} applied to the vector distribution functions.
The corresponding convergence histories for this 3D MHD flow problem
are shown in Fig.~\ref{fig:3Ddrivencavity_Reserror_MHD}.
Again, a significant reduction in the number of time steps to reach
steady state is achieved through preconditioning. Moreover, to put
things in perspective, when the SRT-LBE was employed for the same
grid resolution as above, the simulations were found to be stable
only for $\nu\geq 0.166$. On the other hand, with GLBE, as indicated
above, we could use a much lower value for $0.015$ while maintaining
numerical stability. Thus, for this problem, by using the
preconditioned GLBE rather than the preconditioned SRT-LBE, the
numerical stability is enhanced by almost an order of magnitude.

\begin{figure}
\includegraphics[width = 105mm]{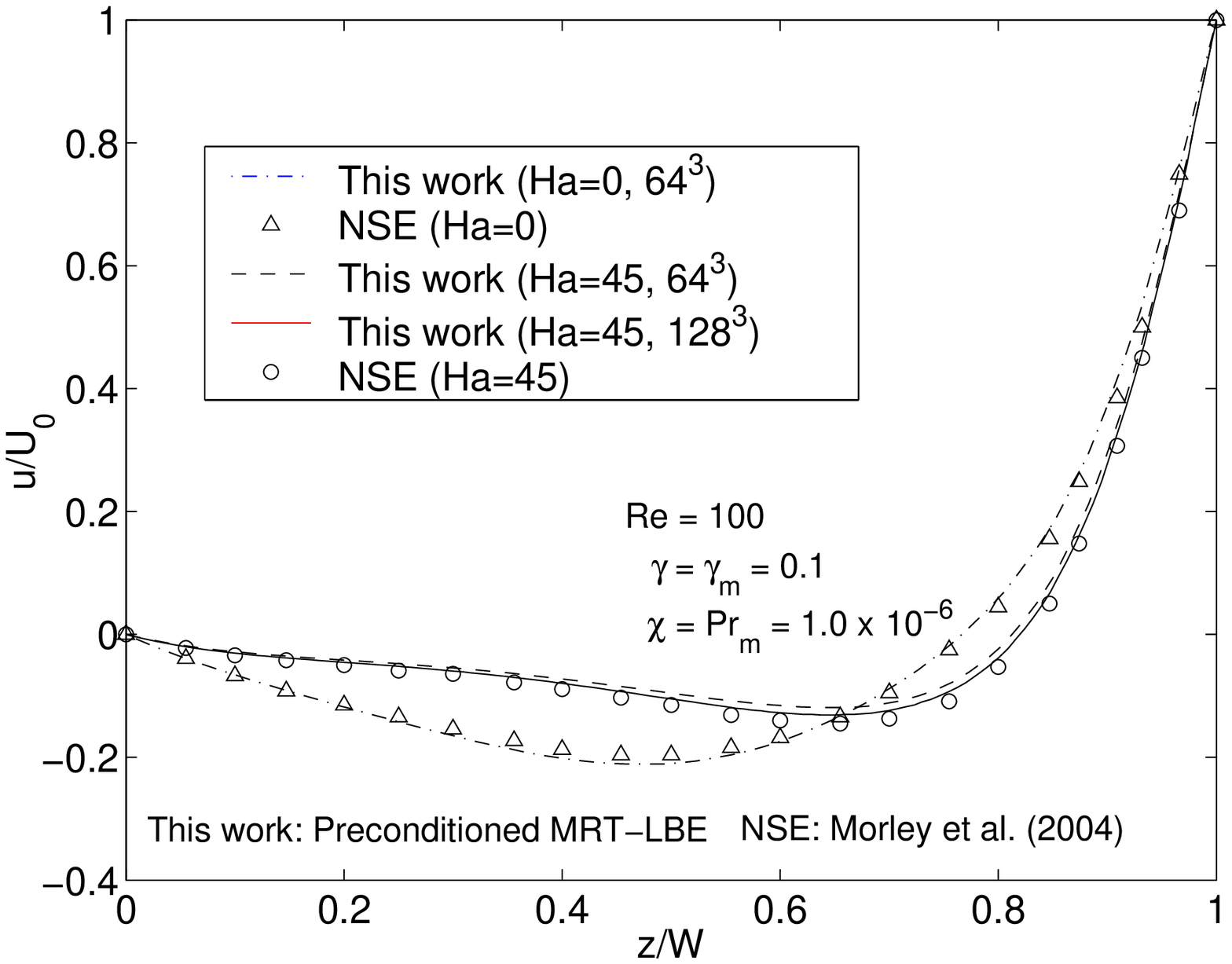}
\caption{\label{fig:drivencavity_uz} Comparison of computed velocity
profile $u$ along the line $x=W/2$ and $y=W/2$ using the
preconditioned MRT-LBE with finite-difference solution of NSE
(Morley et al. (2004)) for simulation of 3D cubical lid-driven
cavity flow with and without magnetic field, i.e., $\Ha=0$ and
$\Ha=45$, respectively; $\Reyn=100$.}
\end{figure}

\begin{figure}
\includegraphics[width = 105mm]{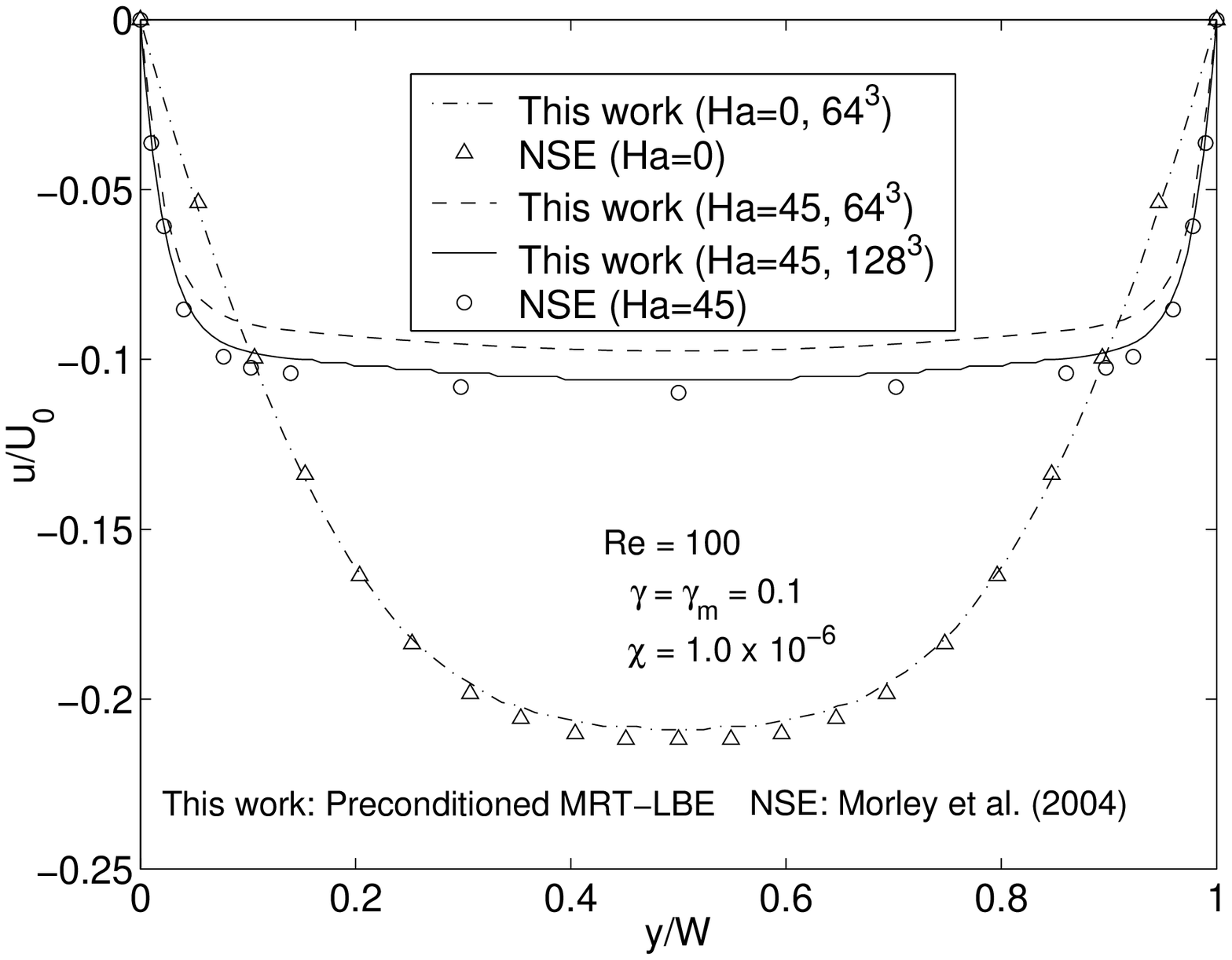}
\caption{\label{fig:drivencavity_uy} Comparison of computed velocity
profile $u$ along the line $x=W/2$ and $z=W/2$ using the
preconditioned MRT-LBE with finite-difference solution of NSE
(Morley et al. (2004)) for simulation of 3D cubical lid-driven
cavity flow with and without magnetic field, i.e., $\Ha=0$ and
$\Ha=45$, respectively; $\Reyn=100$.}
\end{figure}

\begin{figure}
\includegraphics[width = 105mm]{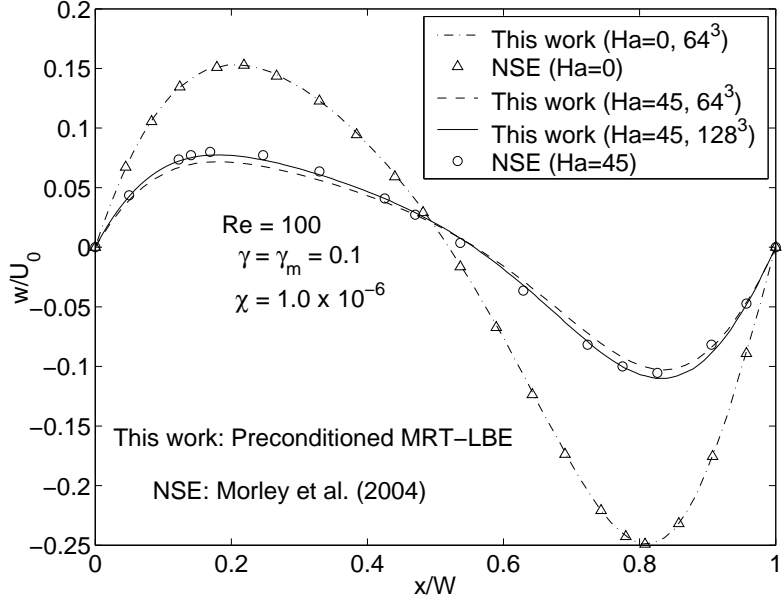}
\caption{\label{fig:drivencavity_wx} Comparison of computed velocity
profile $w$ along the line $y=W/2$ and $z=W/2$ using the
preconditioned MRT-LBE with finite-difference solution of NSE
(Morley et al. (2004)) for simulation of 3D cubical lid-driven
cavity flow with and without magnetic field, i.e., $\Ha=0$ and
$\Ha=45$, respectively; $Re=100$.}
\end{figure}

We will now investigate the accuracy of preconditioned LBE for this
problem.
Figures~\ref{fig:drivencavity_uz},~\ref{fig:drivencavity_uy}
and~\ref{fig:drivencavity_wx} show the computed velocity profiles
for the cases with $\Ha=0$ (i.e. no magnetic field) and $\Ha=45$ and
compared with recent results from simulations carried out using
finite-difference method (FDM) involving the solution of the
Navier-Stokes equations~\cite{morley04}.
The presence of Lorentz forces appears to significantly influence
the characteristics of fluid motion in this 3D problem. In
particular, the velocity profile appears to be markedly flattened by
the presence of magnetic field. The computed results are in
excellent agreement with the FDM. It was noticed that the velocity
profile bounded by the Hartmann walls (i.e. those perpendicular to
the direction of $B_0$, see
Fig.~\ref{fig:3D_Driven_Cavity_Schematic}), is somewhat sensitive to
the grid resolution employed. Morley \emph{et al}.~\cite{morley04},
who studied this problem using different numerical schemes with
different grid resolutions, also observed such effects. In this work,
we find that by further refining the grid, i.e. by doubling the number of grid nodes
in each direction, the computed solution converges to the FDM results.

\begin{figure}
\includegraphics[width = 100mm,viewport=115 100 590
500,clip]{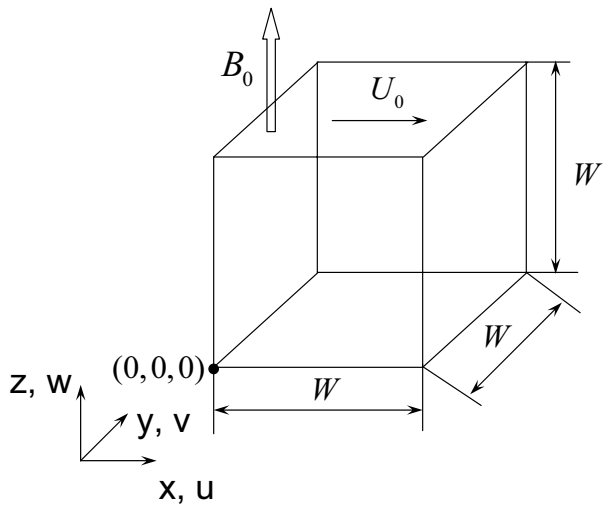}
\caption{\label{fig:3D_Driven_Cavity_Diff_Schematic} Schematic
illustration of a three-dimensional (3D) cubical cavity with its top
lid moving at velocity $U_0$ in the presence of an external magnetic
field $B_0$, applied perpendicular to both the lid surface and its
direction of motion.}
\end{figure}

\begin{figure}
\includegraphics[width = 105mm]{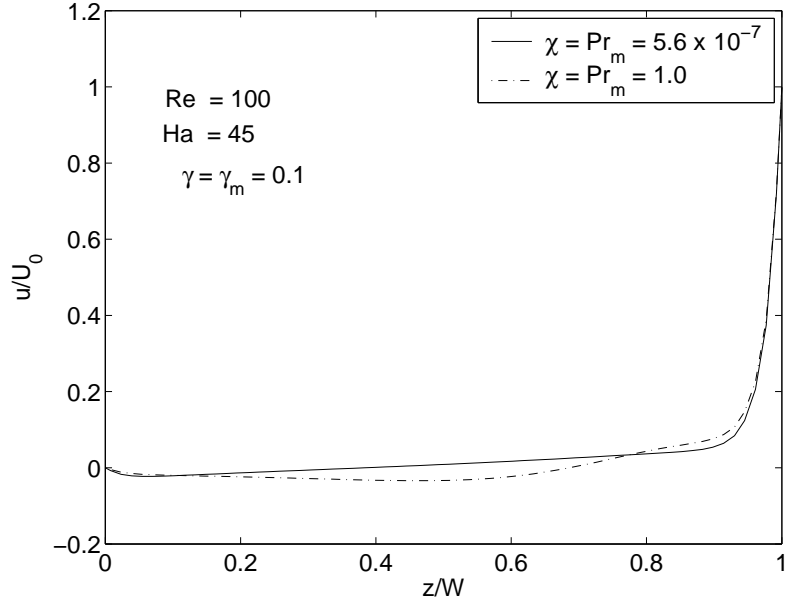}
\caption{\label{fig:drivencavity_lowhighuz} Effect of magnetic
Prandtl number $\Pran_m$ on computed velocity profile $u$ along the
line $x=W/2$ and $y=W/2$ for simulation of 3D cubical lid-driven
cavity in the presence of magnetic field perpendicular to the lid
surface and its direction of motion; $\Reyn=100$ and $\Ha=45$.}
\end{figure}

\begin{figure}
\includegraphics[width = 105mm]{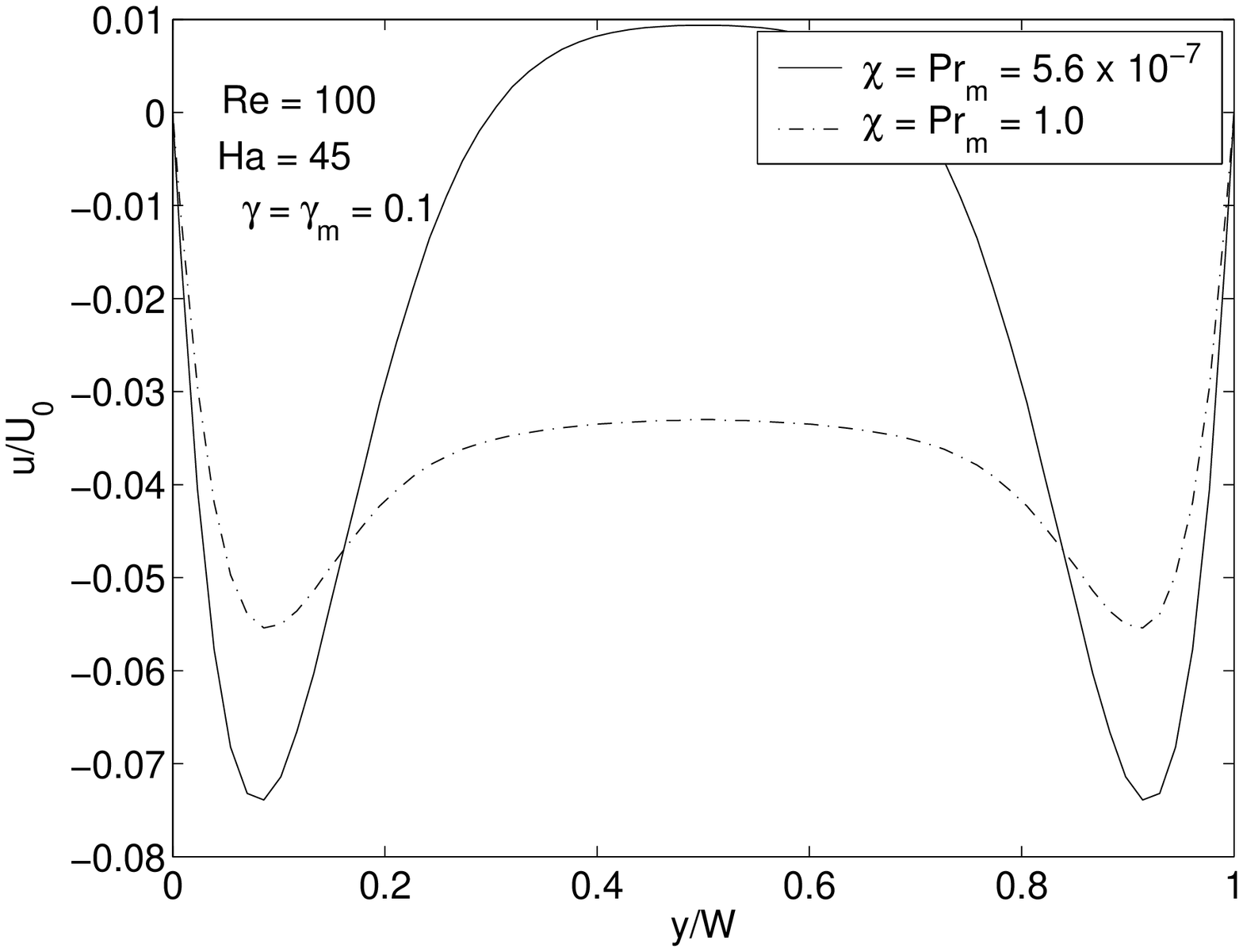}
\caption{\label{fig:drivencavity_lowhighuy} Effect of magnetic
Prandtl number $\Pran_m$ on computed velocity profile $u$ along the
line $x=W/2$ and $z=W/2$ for simulation of 3D cubical lid-driven
cavity in the presence of magnetic field perpendicular to the lid
surface and its direction of motion; $\Reyn=100$ and $\Ha=45$.}
\end{figure}

\begin{figure}
\includegraphics[width = 105mm]{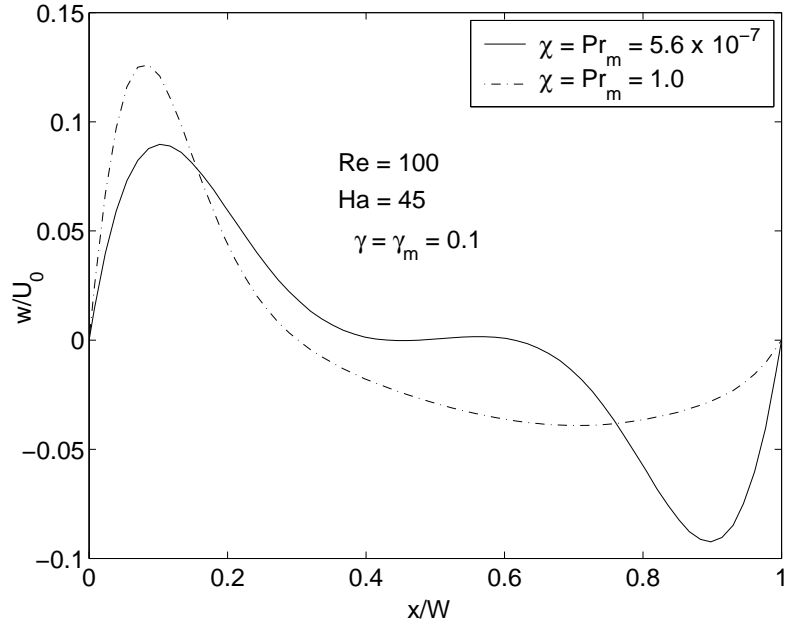}
\caption{\label{fig:drivencavity_lowhighwx} Effect of magnetic
Prandtl number $\Pran_m$ on computed velocity profile $u$ along the
line $y=W/2$ and $z=W/2$ for simulation of 3D cubical lid-driven
cavity in the presence of magnetic field perpendicular to the lid
surface and its direction of motion; $\Reyn=100$ and $\Ha=45$.}
\end{figure}

\begin{figure}
\includegraphics[width = 105mm]{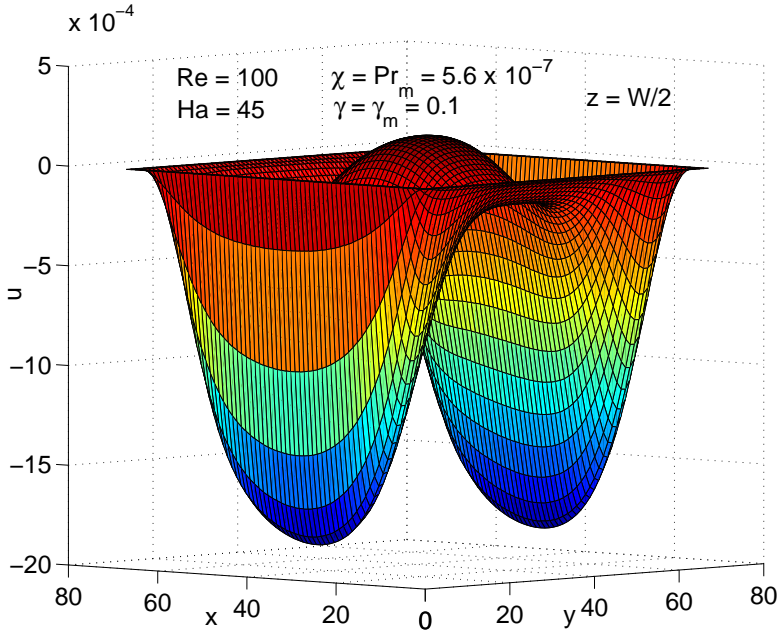}
\caption{\label{fig:lid_wall_jet} Computed velocity field $u$ along
the plane $z=W/2$ showing side wall-jets in a 3D cubical lid-driven
cavity in the presence of magnetic field perpendicular to the lid
surface and its direction of motion; $\Reyn=100$, $\Ha=45$ and $\Pran_m=5.6
\times 10^{-7}$.}
\end{figure}

For the sake of completeness, we will now consider 3D MHD driven
cavity flow with the magnetic field applied in a different manner,
i.e. $B_0$ perpendicular to the lid surface and its direction of
motion as shown in Fig.~\ref{fig:3D_Driven_Cavity_Diff_Schematic}.
\clearpage \pagebreak We will consider the effect of magnetic
Prandtl number $\Pran_m$ for this case by employing different values of
the scaling factor $\chi$ in the preconditioned LBE.
Figures~\ref{fig:drivencavity_lowhighuz},~\ref{fig:drivencavity_lowhighuy}
and~\ref{fig:drivencavity_lowhighwx} show the velocity profiles
along different directions for $\Reyn=100$, $\Ha=45$ with
$\gamma=\gamma_m=0.1$ for two values of $\Pran_m$, i.e.,
$\Pran_m=5.6\times 10^{-7}$ and $\Pran_m=1.0$, with the former
corresponding to liquid metal.
It is noticed that the values of $\Pran_m$ appears to strongly modulate
the velocity field for those cases in which they are bounded on both
sides by stationary walls (Figs.~\ref{fig:drivencavity_lowhighuy}
and~\ref{fig:drivencavity_lowhighwx}). Moreover, the direction of
the application of magnetic field appears to have a profound
influence on the flow field. In particular, in contrast to the
previous case, we notice the presence of wall jet like features when
$B_0$ is perpendicular to both the surface of the top lid and its
direction of motion (see Fig.~\ref{fig:lid_wall_jet}).

\newpage

\section{\label{sec:summary}Summary and Conclusions}
In this paper, we devised a preconditioning approach for
accelerating the steady state convergence of the solution of the
generalized lattice Boltzmann equation (GLBE) with forcing term representing
non-uniform force fields. A preconditioning parameter is introduced in the
equilibrium moments and in the forcing terms of the GLBE that alleviates the disparity
between the propagation speeds of density perturbation and fluid
flow processes at low Mach numbers. The use of multiple relaxation
times in the collision step involving the solution of the GLBE
significantly improves the numerical stability of the approach as compared with
the single relaxation time approach. Particular attention is paid to consistently
preconditioning the projections of the forcing terms in the natural moment space of the
GLBE. In particular, it is shown that to avoid spurious effects due
to preconditioning, the slower processes involving the interaction
of the velocity field and the external forces need to be
preconditioned by the square of the prconditioning parameter. The
limiting form of the consistent forcing term, which avoids discrete
lattice artifacts particularly for spatially and temporally
dependent external forces, is also obtained in the case of SRT-LBE.

As an example of preconditioning an extended system of LBE for
complex flows, we also developed a preconditioning procedure for
a vector lattice kinetic scheme that accelerate convergence of
magnetohydrodynamics (MHD) equations. In the case of MHD flows, there is an
additional disparity between the propagation speeds of the
perturbation of magnetic induction and the perturbation of the
density field, which is mitigated by preconditioning an additional
parameter. Motivated by applications in fusion engineering involving
handling of liquid metals as blanket walls, in the context of
preconditioning, we also devised a simple strategy to simulate flows
with low magnetic Reynolds numbers or low magnetic Prandtl numbers. The
greatest reduction in the number of time steps to reach steady state
is obtained when the preconditioning parameters of both the GLBE
and the lattice kinetic scheme are the same. The preconditioned approach
yielded solutions in very good agreement with prior solutions for several
canonical examples. It is found that the preconditioned LBE reduced the
number of steps for steady state convergence between several factors and as
much as one or two orders of magnitude depending on the problem. In
particular, for 3D MHD cavity flows that are characterized by shear
and other complex flow features, the preconditioning approach also
yields accurate results in comparison with other numerical
solutions. The preconditioning approach presented in this paper can be
readily extended for other complex problems such as multiphase flows.

\section*{\label{acknowledgements}Acknowledgements}
This work was performed under the auspices of the National
Aeronautics and Space Administration (NASA) under Contract
No.~NNL06AA34P and U.S. Department of Energy (DOE) under Grant
No.~DE-FG02-03ER83715.

\newpage

\appendix

\section{\label{app:momentcomponents} Moments, Equilibrium Moments, Moment-projections of Source Terms}
\subsection{D2Q9 Model}
The components of the various elements in the moments are as
follows~\cite{lallemand00}: $\widehat{f}_0 = \rho, \widehat{f}_1= e,
\widehat{f}_2 = e^2, \widehat{f}_3 = j_x,\widehat{f}_4 =
q_x,\widehat{f}_5 = j_y, \widehat{f}_6 = q_y, \widehat{f}_7 =
p_{xx}, \widehat{f}_8 = p_{xy}$. Here, $\rho$ is the density, $e$
and $e^2$ represent kinetic energy that is independent of density
and square of energy, respectively; $j_x$ and $j_y$ are the
components of the momentum, i.e. $j_x = \rho u_x$ and $j_y = \rho
u_y$, $q_x$, and $q_y$ are the components of the energy flux, and
$p_{xx}$ and $p_{xy}$ are the components of the symmetric traceless
viscous stress tensor.

The corresponding components of the equilibrium moments, which are
functions of the conserved moments, i.e. density $\rho$ and momentum
$\overrightarrow{j}$, are as follows~\cite{lallemand00}:

$\widehat{f}_0^{eq} = \rho, \widehat{f}_1^{eq} \equiv
e^{eq}=-2\rho+3\frac{\overrightarrow{j}\cdot\overrightarrow{j}}{\rho},
\widehat{f}_2^{eq} \equiv
e^{2,eq}=\rho-3\frac{\overrightarrow{j}\cdot\overrightarrow{j}}{\rho},
\widehat{f}_3^{eq} = j_x,\widehat{f}_4^{eq} \equiv
q_x^{eq}=-j_x,\widehat{f}_5^{eq} = j_y, \widehat{f}_6^{eq} \equiv
q_y^{eq}=-j_y, \widehat{f}_7^{eq} \equiv p_{xx}^{eq}=
\frac{(j_x^2-j_y^2)}{\rho}, \widehat{f}_8^{eq} \equiv
p_{xy}^{eq}=\frac{j_xj_y}{\rho}$.

The components of the source terms in moment space are functions of
external force $\overrightarrow{F}$ and velocity fields
$\overrightarrow{u}$, and are given as follows: $\widehat{S}_0 = 0,
\widehat{S}_1 = 6(F_xu_x+F_yu_y), \widehat{S}_2 = -6(F_xu_x+F_yu_y),
\widehat{S}_3= F_x, \widehat{S}_4 =-F_x, \widehat{S}_5=F_y,
\widehat{S}_6 =-F_y, \widehat{S}_7=2(F_xu_x-F_yu_y),
\widehat{S}_8=(F_xu_y+F_yu_x)$.

\subsection{D3Q19 Model}
The components of the various elements in the moments are as
follows~\cite{dhumieres02}: $\widehat{f}_0 = \rho, \widehat{f}_1= e,
\widehat{f}_2 = e^2, \widehat{f}_3 = j_x,\widehat{f}_4 =
q_x,\widehat{f}_5 = j_y, \widehat{f}_6 = q_y, \widehat{f}_7 = j_z,
\widehat{f}_8 = q_z, \widehat{f}_9 = 3p_{xx},\widehat{f}_{10} =
3\pi_{xx},\widehat{f}_{11} = p_{ww},\widehat{f}_{12} =
\pi_{ww},\widehat{f}_{13} = p_{xy},\widehat{f}_{14} =
p_{yz},\widehat{f}_{15} = p_{xz},\widehat{f}_{16} =
m_x,\widehat{f}_{17} = m_y,\widehat{f}_{18} = m_z$. Here, $\rho$ is
the density, $e$ and $e^2$ represent kinetic energy that is
independent of density and square of energy, respectively; $j_x$,
$j_y$ and $j_z$ are the components of the momentum, i.e. $j_x = \rho
u_x$, $j_y = \rho u_y$, $j_z = \rho u_z$, $q_x$, $q_y$, $q_z$ are
the components of the energy flux, and $p_{xx}$, $p_{xy}$, $p_{yz}$
and $p_{xz}$ are the components of the symmetric traceless viscous
stress tensor; The other two normal components of the viscous stress
tensor, $p_{yy}$ and $p_{zz}$, can be constructed from $p_{xx}$ and
$p_{ww}$, where $p_{ww} = p_{yy} - p_{zz}$. Other moments include
$\pi_{xx}$, $\pi_{ww}$, $m_x$, $m_y$ and $m_z$. The first two of
these moments have the same symmetry as the diagonal part of the
traceless viscous tensor $p_{ij}$, while the last three vectors are
parts of a third rank tensor, with the symmetry of $j_kp_{mn}$.

The corresponding components of the equilibrium moments, which are
functions of the conserved moments, i.e. density $\rho$ and momentum
$\overrightarrow{j}$, are as follows~\cite{dhumieres02}:

$\widehat{f}_0^{eq} = \rho, \widehat{f}_1^{eq} \equiv
e^{eq}=-11\rho+19\frac{\overrightarrow{j}\cdot\overrightarrow{j}}{\rho},
\widehat{f}_2^{eq} \equiv
e^{2,eq}=3\rho-\frac{11}{2}\frac{\overrightarrow{j}\cdot\overrightarrow{j}}{\rho},
\widehat{f}_3^{eq} = j_x,\widehat{f}_4^{eq} \equiv
q_x^{eq}=-\frac{2}{3}j_x,\widehat{f}_5^{eq} = j_y,
\widehat{f}_6^{eq} \equiv q_y^{eq}=-\frac{2}{3}j_y,
\widehat{f}_7^{eq} = j_z, \widehat{f}_8^{eq} \equiv
q_z^{eq}=-\frac{2}{3}j_z, \widehat{f}_9^{eq} \equiv
3p_{xx}^{eq}=\frac{\left[3j_x^2-\overrightarrow{j}\cdot\overrightarrow{j}
\right]}{\rho},\widehat{f}_{10}^{eq} \equiv
3\pi_{xx}^{eq}=3\left(-\frac{1}{2}p_{xx}^{eq}
\right),\widehat{f}_{11}^{eq} \equiv
p_{ww}^{eq}=\frac{\left[j_y^2-j_z^2
\right]}{\rho},\widehat{f}_{12}^{eq} \equiv
\pi_{ww}^{eq}=-\frac{1}{2}p_{ww}^{eq},\widehat{f}_{13}^{eq} \equiv
p_{xy}^{eq}=\frac{j_xj_y}{\rho},\widehat{f}_{14}^{eq} \equiv
p_{yz}^{eq}=\frac{j_yj_z}{\rho},\widehat{f}_{15}^{eq} \equiv
p_{xz}^{eq}=\frac{j_xj_z}{\rho},\widehat{f}_{16}^{eq} =
0,\widehat{f}_{17}^{eq} = 0,\widehat{f}_{18}^{eq} = 0$.

The components of the source terms in moment space are functions of
external force $\overrightarrow{F}$ and velocity fields
$\overrightarrow{u}$, and are given as follows~\cite{premnath08}:

$\widehat{S}_0 = 0, \widehat{S}_1 = 38(F_xu_x+F_yu_y+F_zu_z),
\widehat{S}_2 = -11(F_xu_x+F_yu_y+F_zu_z), \widehat{S}_3= F_x,
\widehat{S}_4 =-\frac{2}{3}F_x, \widehat{S}_5=F_y, \widehat{S}_6
=-\frac{2}{3}F_y, \widehat{S}_7=F_z, \widehat{S}_8 =
-\frac{2}{3}F_z, \widehat{S}_9 = 2(2F_xu_x-F_yu_y-F_zu_z),
\widehat{S}_{10} = -(2F_xu_x-F_yu_y-F_zu_z), \widehat{S}_{11} =
2(F_yu_y-F_zu_z), \widehat{S}_{12} = -(F_yu_y-F_zu_z),
\widehat{S}_{13} = (F_xu_y+F_yu_x), \widehat{S}_{14} = (F_yu_z+F_zu_y),
\widehat{S}_{15} = (F_xu_z+F_zu_x), \widehat{S}_{16} = 0, \widehat{S}_{17} = 0,
\widehat{S}_{18} = 0$.

\section{\label{app:Chapman_Enskog_pGLBE} Chapman-Enskog Analysis of the Preconditioned GLBE with Forcing Term for D2Q9 Model}
In this section, by employing the Chapman-Enskog multiscale
analysis~\cite{chapman64,premnath06}, we derive the
macroscopic dynamical equations for the preconditioned GLBE with
forcing term corresponding to the D2Q9 model. The analysis of the
preconditioned GLBE with other two- or three-dimensional models can
be carried out in an analogous way. First, we introduce the
expansions
\begin{eqnarray}
 \mathbf{\widehat{f}}&=&\sum_{n=0}^{\infty}\epsilon^n \mathbf{\widehat{f}}^{(n)},\\
\partial_t&=&\sum_{n=0}^{\infty}\epsilon^n \partial_{t_n},
\end{eqnarray}
where $\epsilon=\delta_t$, along with the Taylor series into the
preconditioned GLBE presented in Sec.~\ref{subsec:pglbe}. Then,
recognizing that it was derived after making use of the the
transformation
$\mathbf{\widehat{\overline{f}}}=\mathbf{\widehat{f}}-1/2\mathbf{\widehat{S}}$
on a second-order time discretization of the source terms to make it
effectively explicit, and dropping the ``overbar" subsequently for
convenience, the following equations are obtained as consecutive
orders of the parameter $\epsilon$ in moment space (see
Ref.~\cite{premnath06} for details)
\begin{eqnarray}
O(\epsilon^0): \mathbf{\widehat{f}}^{(0)}&=&\mathbf{\widehat{f}}^{eq,*}\label{eq:morder0mrt3},\\
O(\epsilon^1): \left(\partial_{t_0}+ \widehat{\mathcal{E}}_i
\partial_i \right) \mathbf{\widehat{f}}^{(0)}&=&
                -  \widehat{\Lambda}^{*}  \mathbf{\widehat{f}}^{(1)} +
       \mathbf{\widehat{S}}^{*}\label{eq:morder1mrt3},\\
O(\epsilon^2): \partial_{t_1} \mathbf{\widehat{f}}^{(0)}+
                              \left(\partial_{t_0}+   \widehat{\mathcal{E}}_i \partial_i\right)
                  \left( \mathcal{I}   -\frac{1}{2} \widehat{\Lambda}^{*} \right)
                  \mathbf{\widehat{f}}^{(1)}&=&
                 -  \widehat{\Lambda}^{*} \mathbf{\widehat{f}}^{(2)},
\label{eq:morder2mrt3}
\end{eqnarray}
where $ \widehat{\mathcal{E}}_i =\mathcal{T} e_{\alpha i}
\mathcal{T}^{-1}$. As it is known that the non-linear terms in the
equilibrium moments give rise to slower convective behavior of the
fluid as compared with the propagation speed of the density
perturbations, we precondition those terms by the parameter $\gamma$
to obtain $\mathbf{\widehat{f}}^{eq,*}$. The forcing terms in moment
space are functions of external forces and velocity fields as given
in Appendix~\ref{app:momentcomponents}. We assume that for the
components linear in $\overrightarrow{F}$, we precondition them by
$\gamma$. On the other hand, we precondition those that are non-linear due to
interactions between $\overrightarrow{F}$ and $\overrightarrow{u}$ by an
unknown parameter $\gamma_1$, whose form will be deduced as part of the
analysis. The final expressions for the preconditioned quantities
$\mathbf{\widehat{f}}^{eq,*}$ and $\mathbf{S}^{*}$, as well as
$\widehat{\Lambda}^{*}$ are given in Sec.~\ref{subsec:pglbe}.

The components of the first-order equations in moment space, i.e.
Eq.~(\ref{eq:morder1mrt3}) can be written as
\begin{equation}
\partial_{t_0}\rho+\partial_xj_x+\partial_yj_y=0
\label{eq:fmom0}
\end{equation}
\begin{equation}
\partial_{t_0}\left(-2\rho+\frac{\overrightarrow{j}\cdot\overrightarrow{j}}{\gamma\rho}\right)=-s_1^*e^{(1)}+6\frac{\overrightarrow{F}\cdot\overrightarrow{u}}{\gamma_1}
\label{eq:fmom1}
\end{equation}
\begin{equation}
\partial_{t_0}\left(\rho-3\frac{\overrightarrow{j}\cdot\overrightarrow{j}}{\gamma\rho}\right)-\overrightarrow{\nabla}\cdot\overrightarrow{j}=-s_2^*e^{2(1)}-6\frac{\overrightarrow{F}\cdot\overrightarrow{u}}{\gamma_1}
\label{eq:fmom2}
\end{equation}
\begin{equation}
\partial_{t_0}j_x+\partial_x\left(\frac{1}{3}\rho+\frac{j_x^2}{\gamma\rho}\right)+\partial_y\left(\frac{j_xj_y}{\gamma\rho}\right)=\frac{F_x}{\gamma}
\label{eq:fmom3}
\end{equation}
\begin{equation}
\partial_{t_0}\left(-j_x\right)+\partial_x\left(-\frac{1}{3}\rho-\frac{j_x^2-j_y^2}{\gamma\rho}\right)+\partial_y\left(\frac{j_xj_y}{\gamma\rho}\right)=-s_4^*q_x^{(1)}-\frac{F_x}{\gamma}
\label{eq:fmom4}
\end{equation}
\begin{equation}
\partial_{t_0}j_y+\partial_x\left(\frac{j_xj_y}{\gamma\rho}\right)+\partial_y\left(\frac{1}{3}\rho+\frac{j_y^2}{\gamma\rho}\right)=\frac{F_y}{\gamma}
\label{eq:fmom5}
\end{equation}
\begin{equation}
\partial_{t_0}\left(-j_y\right)+\partial_x\left(\frac{j_xj_y}{\gamma\rho}\right)+\partial_y\left(-\frac{1}{3}\rho+\frac{j_x^2-j_y^2}{\gamma\rho}\right)=-s_6^*q_y^{(1)}-\frac{F_y}{\gamma}
\label{eq:fmom6}
\end{equation}
\begin{equation}
\partial_{t_0}\left(\frac{j_x^2-j_y^2}{\gamma\rho}\right)+\partial_x\left(\frac{2}{3}j_x\right)-\partial_y\left(\frac{2}{3}j_y\right)=-s_7^*p_{xx}^{(1)}+2\frac{F_xu_x-F_yu_y}{\gamma_1}
\label{eq:fmom7}
\end{equation}
\begin{equation}
\partial_{t_0}\left(\frac{j_xj_y}{\gamma\rho}\right)+\partial_x\left(\frac{1}{3}j_x\right)+\partial_y\left(\frac{1}{3}j_y\right)=-s_8^*p_{xy}^{(1)}+\frac{F_xu_y+F_yu_x}{\gamma_1}
\label{eq:fmom8}
\end{equation}

Similarly, the components of the second-order equations in moment
space, i.e. Eq.~(\ref{eq:morder2mrt3}) are
\begin{equation}
\partial_{t_1}\rho=0
\label{eq:smom0}
\end{equation}
\begin{eqnarray}
&&\partial_{t_1}\left(-2\rho+\frac{\overrightarrow{j}\cdot\overrightarrow{j}}{\gamma\rho}\right)+\partial_{t_0}\left(\left[1-\frac{1}{2}s_1^*\right]
e^{(1)}\right)+\nonumber\\
&&\partial_x\left(\left[1-\frac{1}{2}s_4^*\right]
q_x^{(1)}\right)+\partial_y\left(\left[1-\frac{1}{2}s_6^*\right]
q_y^{(1)}\right)=-s_1^*e^{(2)} \label{eq:smom1}
\end{eqnarray}
\begin{eqnarray}
&&\partial_{t_1}\left(\rho-3\frac{\overrightarrow{j}\cdot\overrightarrow{j}}{\gamma\rho}\right)+\partial_{t_0}\left(\left[1-\frac{1}{2}s_2^*\right]
e^{2(1)}\right)+\nonumber\\
&&\partial_x\left(\left[1-\frac{1}{2}s_4^*\right]
q_x^{(1)}\right)+\partial_y\left(\left[1-\frac{1}{2}s_6^*\right]
q_y^{(1)}\right)=-s_2^*e^{2(2)} \label{eq:smom2}
\end{eqnarray}
\begin{eqnarray}
&&\partial_{t_1}j_x+
\partial_x\left(\frac{1}{6}\left[1-\frac{1}{2}s_1^*\right]
e^{(1)}+\frac{1}{2}\left[1-\frac{1}{2}s_7^*\right]
p_{xx}^{(1)}\right)+\nonumber \\
&&\partial_y\left(\left[1-\frac{1}{2}s_8^*\right]
p_{xy}^{(1)}\right)=0 \label{eq:smom3}
\end{eqnarray}
\begin{eqnarray}
&&\partial_{t_1}\left(-j_x\right)+
\partial_{t_0}\left(\left[1-\frac{1}{2}s_4^*\right]
q_{x}^{(1)}\right)+\nonumber\\
&&
\partial_x\left(\frac{1}{3}\left[1-\frac{1}{2}s_1^*\right]
e^{(1)}+\frac{1}{3}\left[1-\frac{1}{2}s_2^*\right]
e^{2(1)}-\left[1-\frac{1}{2}s_7^*\right]
p_{xx}^{(1)}\right)+\nonumber \\
&&\partial_y\left(\left[1-\frac{1}{2}s_8^*\right]
p_{xy}^{(1)}\right)=-s_4^*q_x^{(2)}\label{eq:smom4}
\end{eqnarray}
\begin{eqnarray}
&&\partial_{t_1}j_y+
\partial_x\left(\left[1-\frac{1}{2}s_8^*\right]
p_{xy}^{(1)}\right)+\nonumber \\
&&\partial_y\left(\frac{1}{6}\left[1-\frac{1}{2}s_1^*\right]
e^{(1)}-\frac{1}{2}\left[1-\frac{1}{2}s_7^*\right]
p_{xx}^{(1)}\right)=0 \label{eq:smom5}
\end{eqnarray}
\begin{eqnarray}
&&\partial_{t_1}\left(-j_y\right)+
\partial_{t_0}\left(\left[1-\frac{1}{2}s_6^*\right]
q_{y}^{(1)}\right)+
\partial_x\left(\left[1-\frac{1}{2}s_8^*\right]
p_{xy}^{(1)}\right)+\nonumber \\
&&\partial_y\left(\frac{1}{3}\left[1-\frac{1}{2}s_1^*\right]
e^{(1)}+\frac{1}{3}\left[1-\frac{1}{2}s_2^*\right]
e^{2(1)}+\left[1-\frac{1}{2}s_7^*\right]
p_{xx}^{(1)}\right)=-s_6^*q_y^{(2)} \label{eq:smom6}
\end{eqnarray}
\begin{eqnarray}
&&\partial_{t_1}\left(\frac{j_x^2-j_y^2}{\gamma\rho}\right)+
\partial_{t_0}\left(\left[1-\frac{1}{2}s_7^*\right]
p_{xx}^{(1)}\right)+\nonumber\\
&&
\partial_x\left(-\frac{1}{3}\left[1-\frac{1}{2}s_4^*\right]
q_{x}^{(1)}\right)+\partial_y\left(\frac{1}{3}\left[1-\frac{1}{2}s_6^*\right]
q_{y}^{(1)}\right)=-s_7^*p_{xx}^{(2)} \label{eq:smom7}
\end{eqnarray}
\begin{eqnarray}
&&\partial_{t_1}\left(\frac{j_xj_y}{\gamma\rho}\right)+
\partial_{t_0}\left(\left[1-\frac{1}{2}s_8^*\right]
p_{xy}^{(1)}\right)+\nonumber\\
&&
\partial_x\left(\frac{1}{3}\left[1-\frac{1}{2}s_6^*\right]
q_{y}^{(1)}\right)+\partial_y\left(\frac{1}{3}\left[1-\frac{1}{2}s_4^*\right]
q_{x}^{(1)}\right)=-s_8^*p_{xy}^{(2)} \label{eq:smom8}
\end{eqnarray}

To obtain preconditioned hydrodynamical equations, we need to
combine the evolution equations of moments corresponding to
Eqs.~(\ref{eq:fmom0}),~(\ref{eq:fmom3}) and~(\ref{eq:fmom5}) with
Eqs.~(\ref{eq:smom0}),~(\ref{eq:smom3}) and~(\ref{eq:smom5}),
respectively. Inspection of these six equations show that we need
explicit expressions for the following non-equilibrium moments:
$e^{(1)}$, $p_{xx}^{(1)}$ and $p_{xy}^{(1)}$. Thus, from
Eqs.~(\ref{eq:fmom1}),~(\ref{eq:fmom7}) and~(\ref{eq:fmom8}), we get

\begin{equation}
e^{(1)}=\frac{1}{s_1^*}\left[-\partial_{t_0}\left(-2\rho+3\frac{\overrightarrow{j}\cdot\overrightarrow{j}}{\gamma\rho}\right)+6\frac{\overrightarrow{F}\cdot\overrightarrow{u}}{\gamma_1}\right]
\label{eq:e1}
\end{equation}

\begin{equation}
p_{xx}^{(1)}=\frac{1}{s_6^*}\left[-\partial_{t_0}\left(\frac{j_x^2-j_y^2}{\gamma\rho}\right)-\partial_x\left(\frac{2}{3}j_x\right)+\partial_y\left(\frac{2}{3}j_y\right)+2\frac{\left(F_xu_x-F_yu_y\right)}{\gamma_1}\right]
\label{eq:pxx1}
\end{equation}

\begin{equation}
p_{xy}^{(1)}=\frac{1}{s_7^*}\left[-\partial_{t_0}\left(\frac{j_xj_y}{\gamma\rho}\right)-\partial_x\left(\frac{1}{3}j_x\right)-\partial_y\left(\frac{1}{3}j_y\right)+\frac{\left(F_xu_y+F_yu_x\right)}{\gamma_1}\right]
\label{eq:pxy1}
\end{equation}

Eqs.~(\ref{eq:e1})-(\ref{eq:pxy1}) require time derivatives of the
density and velocity (or momentum) fields, which can be obtained
from Eqs.~(\ref{eq:fmom0}),~(\ref{eq:fmom3}) and~(\ref{eq:fmom5})
and truncating terms of $O(\Ma^2)$ or higher. Thus, we have

$\partial_{t_0}\left( \frac{j_x^2}{\rho} \right)\approx
2\frac{F_xu_x}{\gamma},\qquad \partial_{t_0}\left(
\frac{j_y^2}{\rho} \right)\approx 2\frac{F_yu_y}{\gamma},\qquad
\partial_{t_0}\left( \frac{j_xj_y}{\rho} \right)\approx
2\frac{\left(F_xu_y+F_yu_x\right)}{\gamma}$

Substituting the above relations in Eq.~(\ref{eq:e1})
\begin{equation}
e^{(1)}=\frac{1}{s_1^*}\left[-2\overrightarrow{\nabla}\cdot\overrightarrow{j}-3\left(\frac{2\frac{F_xu_x}{\gamma}+2\frac{F_yu_y}{\gamma}}{\gamma}\right)+6\frac{\overrightarrow{F}\cdot\overrightarrow{u}}{\gamma_1}\right]
\label{eq:e1revised}
\end{equation}
In order to eliminate the dependence of forcing terms in the above
equation, Eq.~(\ref{eq:e1revised}), so that the macrodynamical
equations recover correct physics without any spurious effects, we
need to set
\begin{equation}
\gamma_1=\gamma^2 \label{eq:precondorderrelate}.
\end{equation}
Thus, Eq.~(\ref{eq:precondorderrelate}) suggests that the moment
projections of forcing terms involving non-linear interactions of
external force and velocity fields should be preconditioned by a
factor of $1/\gamma^2$, as they represent slower physical processes
than fluid flow itself. Hence, we get
\begin{equation}
e^{(1)}=-2\frac{1}{s_1^*}\overrightarrow{\nabla}\cdot\overrightarrow{j}.
\label{eq:e1revised1}
\end{equation}
Similarly,
\begin{equation}
p_{xx}^{(1)}=-\frac{2}{3}\frac{1}{s_7^*}\left(\partial_xj_x-\partial_yj_y\right),
\label{eq:pxx1revised1}
\end{equation}
and
\begin{equation}
p_{xy}^{(1)}=-\frac{1}{3}\frac{1}{s_8^*}\left(\partial_xj_y+\partial_yj_x\right).
\label{eq:pxy1revised1}
\end{equation}

So, the preconditioned dynamical equations of the conserved moments
are finally obtained by adding
Eqs.~(\ref{eq:fmom0}),~(\ref{eq:fmom3}) and~(\ref{eq:fmom5}) to
Eqs.~(\ref{eq:smom0}),~(\ref{eq:smom3}) and~(\ref{eq:smom5}),
respectively, after multiplying the latter with $\delta_t$, and
using Eqs.~(\ref{eq:e1revised1})-(\ref{eq:pxy1revised1}). They
correspond to the following preconditioned weakly compressible
Navier-Stokes equations
\begin{equation}
\partial_t\rho+\partial_xj_x+\partial_yj_y=0
\end{equation}
\begin{eqnarray}
&&\partial_tj_x+\frac{1}{\gamma}\left[\partial_x\left(\frac{j_x^2}{\rho}\right)+\partial_y\left(\frac{j_xj_y}{\rho}\right)\right]=-\frac{1}{\gamma}\partial_x
p + \nonumber \\
&&
\frac{1}{\gamma}\partial_x\left(2\nu\left[\partial_xj_x-1/3\overrightarrow{\nabla}\cdot\overrightarrow{j}
\right]+\zeta \overrightarrow{\nabla}\cdot\overrightarrow{j}\right)+
\nonumber \\
&&
\frac{1}{\gamma}\partial_y\left(\nu\left[\partial_xj_y+\partial_yj_x
\right]\right)+\frac{F_x}{\gamma}
\end{eqnarray}
\begin{eqnarray}
&&\partial_tj_y+\frac{1}{\gamma}\left[\partial_x\left(\frac{j_xj_y}{\rho}\right)+\partial_y\left(\frac{j_y^2}{\rho}\right)\right]=-\frac{1}{\gamma}\partial_y
p + \nonumber \\
&&
\frac{1}{\gamma}\partial_x\left(\nu\left[\partial_xj_y+\partial_yj_x
\right]\right)+\nonumber\\
&&
\frac{1}{\gamma}\partial_y\left(2\nu\left[\partial_yj_y-1/3\overrightarrow{\nabla}\cdot\overrightarrow{j}
\right]+\zeta \overrightarrow{\nabla}\cdot\overrightarrow{j}\right)+
\frac{F_y}{\gamma}
\end{eqnarray}
where the pressure field $p$ is given by
\begin{equation}
p=\gamma\frac{1}{3}\rho
\end{equation}
and the transport coefficients, viz., the bulk and shear
viscosities, respectively, as
\begin{equation}
\zeta=\gamma\frac{1}{3}\left[
\frac{1}{s_{1}^*}-\frac{1}{2}\right]\delta_t
\end{equation}
and
\begin{equation}
\nu=\gamma\frac{1}{3}\left[
\frac{1}{s_{\beta}^*}-\frac{1}{2}\right]\delta_t, \quad \beta=7,8
\end{equation}

\section{\label{app:Chapman_Enskog_vectorLBE} Chapman-Enskog Analysis of the Preconditioned Vector Kinetic Equation}
We now perform the Chapman-Enskog analysis of the preconditioned
vector kinetic equation by introducing the expansions
\begin{eqnarray}
g_{\alpha i}&=&\sum_{n=0}^{\infty}\epsilon^n g_{\alpha i}^{(n)},\\
\partial_t&=&\sum_{n=0}^{\infty}\epsilon^n \partial_{t_n},
\end{eqnarray}
with $\epsilon=\delta_t$, in conjunction with the Taylor
series~\cite{chapman64,dellar02}, which result in the following:
\begin{eqnarray}
O(\epsilon^0): g_{\alpha i}^{(0)}&=&g_{\alpha i}^{eq,*}\label{eq:gmorder0mrt3},\\
O(\epsilon^1): \left(\partial_{t_0}+ e_{\alpha j}
\partial_j \right) g_{\alpha i}^{(0)}&=&
                -  \frac{1}{\tau_m^*}  g_{\alpha i}^{(1)} \label{eq:gmorder1mrt3},\\
O(\epsilon^2):
\partial_{t_1} g_{\alpha i}^{(0)}+\left(1-\frac{1}{2\tau_m^*}\right)
\left(\partial_{t_0}+e_{\alpha j} \partial_j\right) g_{\alpha
i}^{(1)} &=& -\frac{1}{\tau_m^*} g_{\alpha i}^{(2)}.
\label{eq:gmorder2mrt3}
\end{eqnarray}
Now, using the following summational constraints
$\sum_{\alpha=0}^{b_m}g_{\alpha j}^{(0)}=B_j$,
$\sum_{\alpha=0}^{b_m}e_{\alpha i}g_{\alpha
j}^{(0)}=\Lambda_{ij}^{(0)}$, $\sum_{\alpha=0}^{b_m}g_{\alpha
j}^{(n)}=0$ and $\sum_{\alpha=0}^{b_m}e_{\alpha i}g_{\alpha
j}^{(n)}=\Lambda_{ij}^{(n)}$ for $n \geq 1$, with
$\Lambda_{ij}^{(0)}=\frac{u_iB_j-B_iu_j}{\gamma_m}$, and taking
zeroth moments of Eqs.~(\ref{eq:gmorder1mrt3}) and
(\ref{eq:gmorder2mrt3}), we get
\begin{eqnarray}
\partial_{t_0}B_i+
\partial_j \Lambda_{ji}^{(0)}&=& 0 \label{eq:gorder1},\\
\partial_{t_1}B_i+\left(1-\frac{1}{2\tau_m^*}\right)
\partial_j \Lambda_{ji}^{(1)} &=& 0.
\label{eq:gorder2}
\end{eqnarray}

Taking the first moment,~i.e. $\sum_{\alpha =0}^{b_m}e_{\alpha
j}(\cdot)$ of Eq.~(\ref{eq:gmorder1mrt3}) and using the identity
$\sum_{\alpha=0}^{b_m}e_{\alpha j}e_{\alpha k}g_{\alpha
i}^{(0))}=\theta_m\delta_{jk}B_i$, where $\delta_{jk}$ is the
Kronecker delta, we get
\begin{equation}
\Lambda_{ji}^{(1)}=-\tau_m^*\left[\partial_{t_0}\Lambda_{ji}^{(0)}+\theta_m\partial_jB_i\right].
\end{equation}
With the scaling $O(B_i)\sim O(u_i)$~\cite{dellar02} and considering
the zeroth-order momentum and magnetic induction equations,
$\partial_{t_0}\Lambda_{ji}^{(0)}\sim O(\Ma^3)$ and hence can be neglected.
As a result,
\begin{equation}
\Lambda_{ji}^{(1)}\approx-\tau_m^*\theta_m\partial_jB_i.
\label{eq:msecondmoment}
\end{equation}

Finally, adding Eq.~(\ref{eq:gorder1}) and
Eq.~(\ref{eq:gorder2})$\times\delta_t$ and using $\partial_t\approx
\partial_{t_0}+\delta_t\partial_{t_1}$, along with
Eq.~(\ref{eq:msecondmoment}), we get the preconditioned magnetic
induction equation
\begin{equation}
\partial_t
B_i+\frac{1}{\gamma_m}\nabla_j\left(u_jB_i-B_ju_i\right)=\frac{1}{\gamma_m}\partial_j\left(
\eta
\partial_j B_i \right),
\label{eq:magneticinduction}
\end{equation}
where $\eta=\gamma_m\theta_m\left(\tau_m^*-1/2\right)\delta_t$.

\end{document}